\documentclass[aps, prl, twocolumn]{revtex4-1}
\usepackage[english]{babel}
\makeatletter
\def\bbl@set@language#1{%
  \edef\languagename{%
    \ifnum\escapechar=\expandafter`\string#1\@empty
    \else\string#1\@empty\fi}%
  \@ifundefined{babel@language@alias@\languagename}{}{%
    \edef\languagename{\@nameuse{babel@language@alias@\languagename}}%
  }%
  \select@language{\languagename}%
  \expandafter\ifx\csname date\languagename\endcsname\relax\else
    \if@filesw
      \protected@write\@auxout{}{\string\select@language{\languagename}}%
      \bbl@for\bbl@tempa\BabelContentsFiles{%
        \addtocontents{\bbl@tempa}{\xstring\select@language{\languagename}}}%
      \bbl@usehooks{write}{}%
    \fi
  \fi}
\newcommand{\DeclareLanguageAlias}[2]{%
  \global\@namedef{babel@language@alias@#1}{#2}%
}
\makeatother
\DeclareLanguageAlias{en}{english}

\usepackage{amsmath,amssymb}
\usepackage{graphicx}
\usepackage{tikz}
\usetikzlibrary{positioning}
\usepackage{color}
\usepackage{siunitx}
\newcommand{\meff}{m_\mathrm{eff}}
\newcommand{\cc}{_\mathrm{cc}}
\newcommand{\bn}{_\mathrm{bn}}
\newcommand{\bp}{_\mathrm{bp}}
\newcommand{\wcc}{w\cc}
\newcommand{\mtarget}{m}
\newcommand{\first}{}

\begin{document}
\title{Tailored ensembles of neural networks optimize sensitivity to stimulus statistics}
\author{Johannes Zierenberg$^{1,2}$}
\author{Jens Wilting$^{1}$}
\author{Viola Priesemann$^{1,2}$}
\author{Anna Levina$^{3,4}$}
\affiliation{
  $^1$ Max Planck Institute for Dynamics and Self-Organization, Am Fassberg 17, 37077 G{\"o}ttingen, Germany,\\
  $^2$ Bernstein Center for Computational Neuroscience, Am Fassberg 17, 37077 G{\"o}ttingen, Germany,\\
  $^3$ University of T\"ubingen, Max Planck Ring 8, 72076 T\"ubingen, Germany,\\
  $^4$ \mbox{Max Planck Institute for Biological Cybernetics, Max Planck Ring 8, 72076 T\"ubingen, Germany}
}

\begin{abstract}

  The dynamic range of stimulus processing in living organisms is much larger
  than a single neural network can explain. For a generic, tunable spiking
  network we derive that while the dynamic range is maximal at criticality, the
  interval of discriminable intensities is very similar for any network tuning
  due to coalescence. Compensating coalescence enables adaptation of
  discriminable intervals. Thus, we can tailor an ensemble of networks optimized
  to the distribution of stimulus intensities, e.g., extending the dynamic range
  arbitrarily. We discuss potential applications in machine learning.

\end{abstract}

\maketitle
\first{Living organisms are constantly exposed to sensory stimuli with
intensities that cover multiple orders of
magnitude~\cite{hecht1924,borg1967,viemeister1988}.} The organisms' ability to
cope with this variability of stimulus intensities determines how well they will
thrive. Hence, evolution favored nervous systems that developed the capability
to process this variability. Stimulus intensity is typically encoded in neural
firing rates, with stronger intensities eliciting higher firing rates. Such reliable,
monotonous encoding of stimulus intensity has for example been found in the
mammalian auditory system~\cite{rouiller1983, viemeister1983, schreiner1992,
dean2005}, with indication that a small number of neurons is sufficient to
discriminate small changes in sound level intensity~\cite{viemeister1983} and
that neurons adjust their response to the complex statistics of the sound level
distribution~\cite{dean2005}.

\first{The capability to process a broad distribution of stimulus intensities
is typically quantified by the dynamic range.} The dynamic range of a neural
network is defined as the log-ratio between the strongest and weakest stimulus
intensities that are reliably encoded by the neural firing rate. Experimentally,
the range of reliably encoded stimulus intensities was measured to extend from
$\SI{40}{dB}$ to $\SI{50}{dB}$ in cat primary auditory nerve
fibers~\cite{evans1980}. This clearly does not cover the full range of hearing
--- ranging for humans from approximately $\SI{0}{dB}$ to about $\SI{120}{dB}$
sound pressure level --- resulting in the so called `dynamic range
problem'~\cite{evans1981, viemeister1988}. It was argued that the dynamic range
problem can be overcome by adaptation of neural response to the presented
stimulus statistics~\cite{dean2005}. However, as the future stimulus intensity
cannot be perfectly predicted, maximizing the dynamic range remains a powerful
strategy. 

\first{The dynamic range of a recurrent neural network, driven by external
stimuli, is maximal at the critical point of the non-driven
network~\cite{kinouchi2006}}. Thereby criticality fosters flexible information
processing, and is thus a versatile candidate state for neural networks and
brain function~\cite{beggs2008, munoz2018}. If already at criticality, the dynamic range
was shown to depend on topology~\cite{wu2007, larremore2011}, with homogeneous
yet random networks reaching a higher dynamic range than those with a
heterogeneous topology. The theoretical results are supported by experiments on
cultured cortex slices~\cite{shew2009}, showing maximal dynamic ranges in those
networks with power-law avalanche-size distributions (signature of criticality).
However, the dynamic range of $\mathcal{O}(\SI{10}{dB})$ observed in cultured
neural networks is more than ten time smaller than the perceptual dynamic range
of humans, and thus recurrent networks alone probably cannot solve the dynamic
range problem. 

\first{In this letter, we show that processing capability might not be
optimized by a single network, but instead by an ensemble of specialized neural
networks that is tuned to stimulus statistics.} First, we note that for an
optimal encoding of stimulus statistics it is not sufficient to maximize the
dynamic range; more importantly, the interval of stimulus intensities that are
reliably discriminated by the network --- the \textit{discriminable interval} --
also needs to cover the relevant intensities. Second, we show analytically that for a
single neural network near criticality, this discriminable interval cannot be
changed, i.e., it cannot be tuned to the relevant stimulus intensities. Third, we
derive how to construct neural networks with adaptive synaptic weights such that
the discriminable interval gradually changes when changing the distance to
criticality, thereby providing a mechanism to tune the network to specific
stimulus intensities.  Last, we demonstrate that a tailored ensemble of such
networks can optimize the ensemble response to stimulus statistics; if needed
with a drastically increased dynamic range that emerges synergistically in the
ensemble.

\begin{figure*}
  \begin{tikzpicture}
    \def\x{4.70cm}
    \def\y{3.3cm}
    \def\dx{-3.8cm}
    \def\dy{ 3.2cm}
    \node at (-\x,+\y){\includegraphics{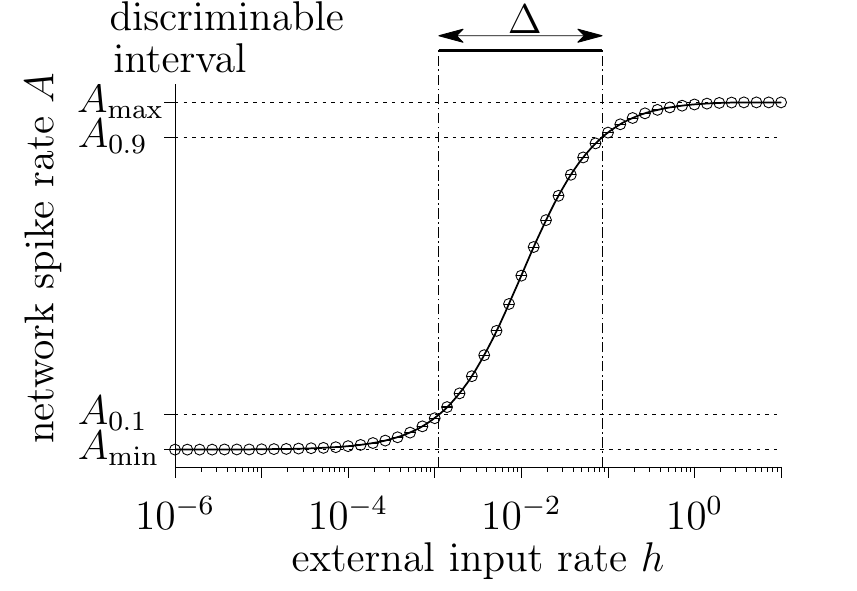}};
    \node at (+\x,+\y){\includegraphics{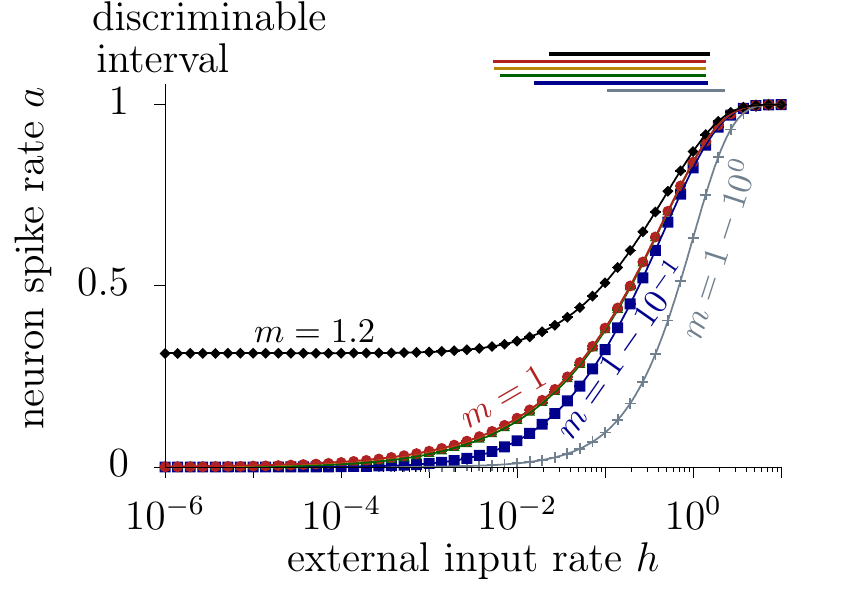}};
    \node at (-\x,-\y){\includegraphics{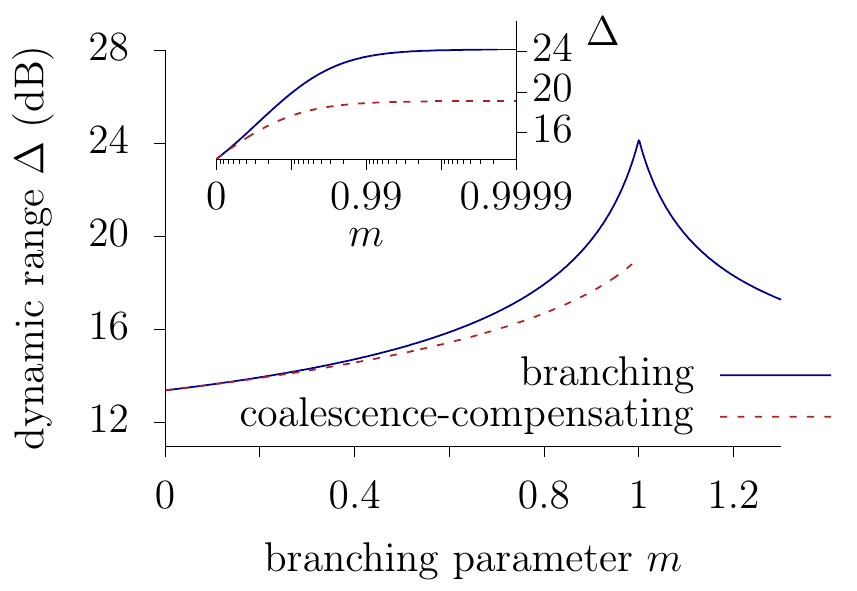}};
    \node at (+\x,-\y){\includegraphics{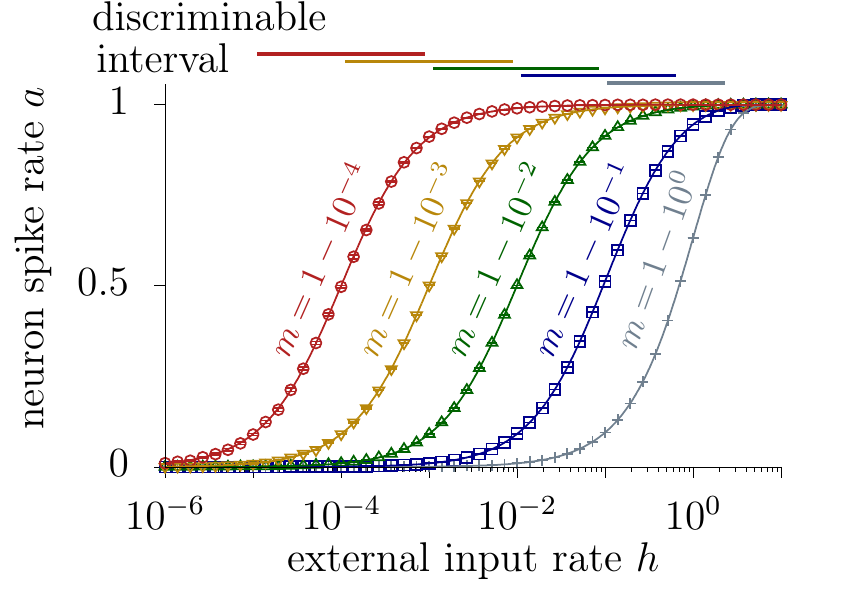}};
    \node at (0cm-\x+\dx,0cm+\y+\dy){$\textbf{A}$};
    \node at (0cm+\x+\dx,0cm+\y+\dy){$\textbf{B}$};
    \node at (0cm-\x+\dx,0cm-\y+\dy){$\textbf{C}$};
    \node at (0cm+\x+\dx,0cm-\y+\dy){$\textbf{D}$};
  \end{tikzpicture}
  \caption{%
    Dynamic range and discriminable interval. 
    \textbf{A} Typical network spike rate as a function of external input rate
    with definition of dynamic range $\Delta$ as the width of the discriminable
    interval.
    \textbf{B} Neuron spike rates and discriminable intervals of the branching network for different branching parameter $m$ ($N=10^4$).
    \textbf{C} Dynamic range as a function of branching parameter for the
    branching network and the coalescence-compensating model. The inset shows
    the dynamic range as a function of distance to the critical point ($m=1$) in
    log scale.
    \textbf{D} Neuron spike rates and discriminable intervals of the
    coalescence-compensating model for different branching parameter $m$ ($N=10^2$).
    \label{figOverview}
    }
\end{figure*}
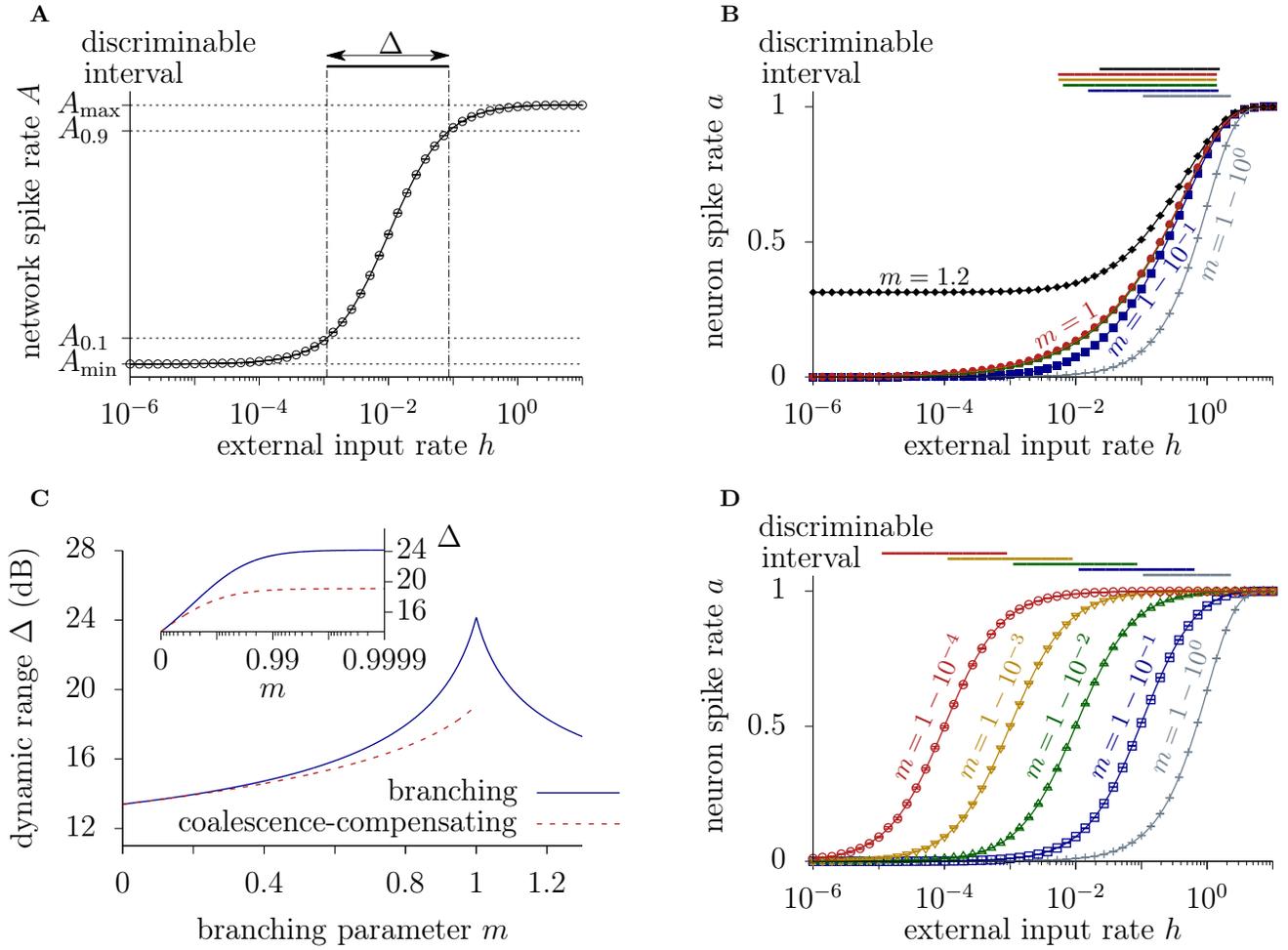
\first{No matter what type of system one analyzes, the dynamic range and the
discriminable interval are response measures defined in terms of the range of
stimulus intensities that can be reliably discriminated in the systems' output
(Fig.~\ref{figOverview}~\textbf{A})}. For a neural network, we treat the
stimulus intensity as an external Poisson input with rate $h$ per neuron,
resulting in a network spike rate $A(h)$. For zero input rate, the network
typically produces its minimal, baseline rate $A_\mathrm{min}$, whereas for very
strong input rates the network rate typically saturates. Hence, the network rate
covers the interval $[A_\mathrm{min},A_\mathrm{max}]$. On this interval, the
discriminable interval is defined as the 10\textsuperscript{th} to
90\textsuperscript{th} percentiles, $\left[h(A_{0.1}),h(A_{0.9})\right]$, where
$A_x=A_\mathrm{min}+x(A_\mathrm{max}-A_\mathrm{min})$. Based on the interval of
discriminable input rates $[h(A_{0.1}),h(A_{0.9})]$, the dynamic range is
defined as $\Delta=10\log_{10}\left[h(A_{0.9})/h(A_{0.1})\right]$ in
decibel~(\SI{}{dB}). These response measures characterize the range of input
rates that can be discriminated in contrast to response measures that
characterize how well a system can detect changes in the input
rate~\cite{chevallier2018}.

\first{We analyze neural network responses using a branching network, because it
shows a maximal dynamic range at
criticality~\cite{haldeman2005,kinouchi2006,zierenberg2018}.} For analytical
tractability, we consider a fully connected network of $N$ binary neurons
without refractory period~\cite{zierenbergLong}. After each time step $\Delta
t$, each neuron $i$ can be either silent ($s^i=0$) or excited to spike
($s^i=1$). It can be excited by (1) an external Poisson input with rate $h$,
such that the transition probability is $\lambda(h)=1-\exp(-h\Delta t)$; or by
(2) a presynaptic neuron $j$ (which was excited in the previous time step) with
transition probability $w^{ij}=w=\frac{m}{N}$, where the branching parameter $m$
is the control parameter. If no external or internal input reaches an excited
neuron, it returns to a silent state in the next time step. The network activity
at each time step, $A_t=\sum_i s^i_{t}$, is thus determined by both internal
($m$) and external ($h$) activation.


\first{For the branching network, we can analytically derive the neuron spike
rate as a function of external input rate.} In the following, we sketch the
main steps and refer to Ref.~\cite{zierenbergLong} for a detailed derivation.
Given a network activity $A_t$ at time $t$, the probability for any neuron to
be excited in the next time step is given by $P\left[s^i_{t+1}=1|
A_t,w,h\right]=1-\left(1-w\right)^{A_t}\left(1-\lambda(h)\right)=p(A_t)$. The
network activity $A_{t+1}$ is then binomially distributed with expectation
value $\langle A_{t+1}|A_t\rangle=Np(A_t)$. Demanding stationary activity,
$A=\langle A_t\rangle=\langle\langle A_{t+1}|A_t\rangle\rangle$, and neglecting
fluctuations in a mean-field approximation, $\langle(1-m/N)^{A_t}\rangle\approx (1-m/N)^{\langle A_t\rangle}$, implies
\begin{equation}\label{eqMFapproximation}
  A=N-N\left(1-w\right)^{A}\left(1-\lambda(h)\right).
\end{equation}  
Using the Lambert-W function defined as $W(z)e^{W(z)}=z$~\cite{corless1996},
this is solved by
{$A(h)=N-W\left[(1-w)^N\ln(1-w)N(1-\lambda(h))\right]/\ln(1-w)$}.  For $w=m/N$
and $N\to\infty$, we can identify $(1-m/N)^N\to e^{-m}$ and expand
$\ln(1-m/N)N=-m+\mathcal{O}\left(N^{-1}\right)$, to obtain the solution for the
expected neuron spike rate 
\begin{equation}\label{eqResponseSTD}
    a\bn(m,h) = \frac{A(h)}{N}=1+\frac{W\left[-m e^{-m}(1-\lambda(h))\right]}{m},
\end{equation}
which turns out to be system-size independent for sufficiently large $N$.
Figure~\ref{figOverview}~\textbf{B} shows that Eq.~\eqref{eqResponseSTD}
accurately describes our numerical results for $N=10^4$.

\first{From the neuron rate, we readily obtain mean-field solutions for the
dynamic range and discriminable interval}. Inverting Eq.~\eqref{eqResponseSTD}
yields 
\begin{equation}\label{eqInvResponseSTD}
  h\bn(m,a) = -\frac{1}{\Delta t}\ln\left[(1-a)e^{ma}\right],
\end{equation}
which allows us to compute $h(a_{0.1})$ and $h(a_{0.9})$. In correspondence with
previous results, we recover that the dynamic range is maximal at criticality
(Fig.~\ref{figOverview}~\textbf{C}). Importantly, it does not diverge for
$\epsilon=1-m\to0$~\cite{wu2007}, which can be seen clearly from the figure
inset.  Moreover, the discriminable interval barely changes for small $\epsilon$
(Fig.~\ref{figOverview}~\textbf{B}). Thus for cortical networks with
$m\approx0.98$~\cite{wilting2018,wilting2018a}, the discriminable intervals of
networks with different $m$ strongly overlap. Strongly overlapping discriminable
intervals are also observed in cultured cortex slices~\cite{shew2009}. 
We can derive the bounds of the discriminable interval $h(a_{x})$ by expanding
$\ln(1-a)=\sum_{n=1}^{\infty} a^n/n$ and rewriting Eq.~\eqref{eqInvResponseSTD}
as $h\bn(m,a)=\left(a\epsilon + \sum_{n=2}^{\infty} a^n/n\right)/\Delta t$ with
$\epsilon=1-m$. For $\epsilon$ sufficiently smaller than $a$, which holds in the
vicinity of the critical point, the bounds are barely distinguishable
(see also Fig.~\ref{figResultDiscriminableInterval}). This result is valid for
sufficiently large system sizes, and limits the dynamic range even in the
infinite system limit. In the following, we will propose a framework that
allows to shift the discriminable interval.

\first{Branching networks are a finite-size network embedding of space-less
branching processes~\cite{harris1963}, which approximate spike propagation in
neuronal tissue~\cite{beggs2003, haldeman2005, kinouchi2006, priesemann2014,
levina2008, levina2007, wilting2018}.} The branching process evolves in discrete
time steps $\Delta t$ and if $A_t$ neurons are excited, one finds on average
$\langle A_{t+1}|A_t\rangle = m A_t$ excited neurons in the next time step.
Considering in addition a network-wide external input at rate $hN$, one expects
$\langle A_{t+1}|A_t\rangle = m A_t +Nh\Delta t$. For stationary activity, $A=
\langle\langle A_{t+1}|A_t\rangle\rangle = mA+Nh\Delta t$, such that
\begin{equation}\label{eqResponseBP}
  a\bp(m,h)=h\Delta t/(1-m).
\end{equation}
The inverse is straightforward, $h\bp(m,a)=(1-m)a/\Delta t$, and leads to a
dynamic range that is independent of $m$ (see Supplemental Material~S3). In
contrast, the discriminable interval is highly dependent on $m$ with bounds
$h\bp(m,a_x) = (1-m)x/\Delta t$. These response measures differ drastically from
those of the branching network.

\first{The response measures of the branching process differ drastically from
those of the branching network, because of \emph{coalescence} (the simultaneous
activation of the same neuron from multiple sources) in the branching
network~\cite{zierenbergLong}.} Coalescence alters the critical non-equilibrium
phase transition from subcritical--supercritical in the branching process to
absorbing--active in the branching network.
We can, however, map the network activity of a branching network to a branching
process with an \emph{effective branching parameter} $\meff(A_t)$ through
$\langle A_{t+1}|A_t\rangle = \meff(A_t)A_t+N\lambda(h)$. Using
Eq.~\eqref{eqMFapproximation}, we can identify
\begin{equation}\label{eqBNconvergence}
  \meff(w,A_t)=
  \left(\frac{N}{A_t}\right)\left(1-\left(1-w\right)^{A_t}\right)\left(1-\lambda(h)\right),
\end{equation}
revealing that coalescence reduces the effective branching parameter
$m_\mathrm{eff}(A_t)<m=wN$.

\first{Having identified coalescence as the main difference, we can compensate
it to construct network dynamics in agreement with the dynamics of a
corresponding branching process}. This can be achieved by inverting
Eq.~\eqref{eqBNconvergence} for a target branching parameter
$\meff(\wcc,A_t)=m$, obtaining adaptive (coalescence-compensating) synaptic
weights $\wcc(A_t)$ (see Supplemental Material~S1). We assume that the network may
at most compensate coalescence with internal sources (setting $\lambda(h)=0$)
such that  
\begin{equation}\label{eqBNcc1}
  \wcc(m,A_t) = \left(1-\left(1-\frac{m A_t}{N}\right)^{1/A_t}\right),
\end{equation}
because compensating coalescence with external input would require knowledge of
the external input rate $h$ (a logical contradiction as one of the network's
tasks is to infer this rate). For finite networks in close vicinity of the
critical point, we need to introduce a finite-network cutoff at
$\wcc(m,N)=\ln(N)/N$, which avoids an additional absorbing boundary at $A=N$
(see Supplemental Material~S1). Compensating internal coalescence with adaptive
weights [Eq.~\eqref{eqBNcc1}] implies 
\begin{equation}\label{eqBNcc2}
  \meff(\wcc(A_t), A_t) = m(1-\lambda(h)),
\end{equation}
such that for small input $h$ we approximate the target branching process
statistics.


\begin{figure}
  \includegraphics{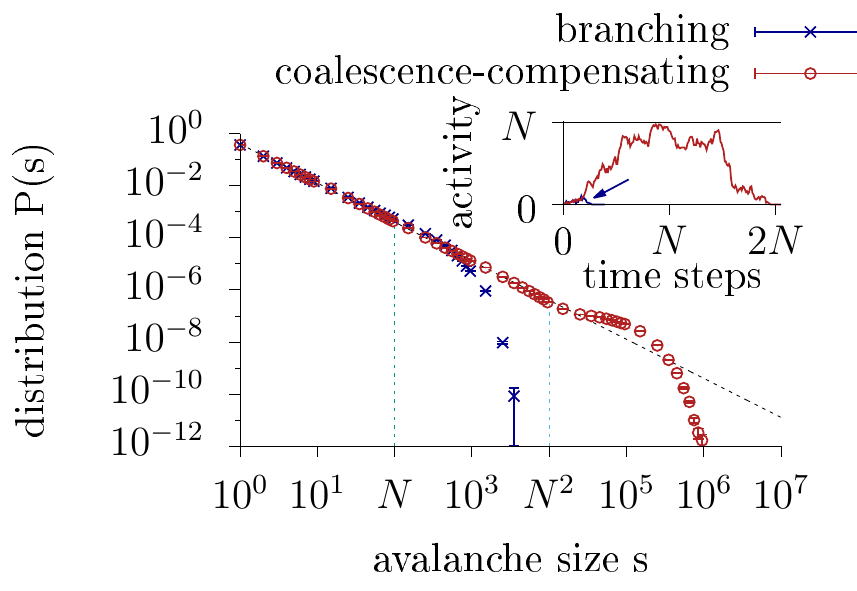}
  \caption{%
    Avalanche-size distribution of a branching and a coalescence-compensating
    network with $N=100$ in the separation of timescale regime, i.e., each
    avalanche is triggered manually ($h\to0$ limit). The inset shows example
    avalanches at the typical upper bounds of the power law distribution
    (vertical dashed line in main plot).
    \label{figCorrections}
  }
\end{figure}
\first{The resulting coalescence-compensating model shows much improved
power-law avalanche-size distributions at criticality
(Fig.~\ref{figCorrections}).} In the branching network the power-law is cut off
at $s\approx N$, for which typical avalanches (Fig.~\ref{figCorrections}, inset)
with bell-shape $A(t,T)=T\mathcal{F}(t/T)$~\cite{friedman2012} imply the
avalanche peak value to scale as $A^\mathrm{peak}\sim T\sim\sqrt{N}$ (see
Supplemental Material~S2). In the coalescence-compensating network, this maximum is
now extended to the non-absorbing boundary $T\sim N$, such that the power-law
characteristics extend until $s\approx N^2$. This means that for the
convergence-compensating network the avalanche-size distribution covers twice as
many orders of magnitude as for the branching network. Similarly, the power-law
avalanche-duration distribution is extended from $\sqrt{N}$ to $N$ (see
Supplemental Material~S2).

\first{The coalescence-compensating network retains maximal dynamic range at
criticality but recovers the dependence of discriminable interval on the
branching parameter $m$.} To show this we compute the neuron rate
for the coalescence-compensating network. For stationary activity,
$A=\langle\langle A_{t+1}|A_t\rangle\rangle=m[1-\lambda(h)] A + N\lambda(h)$,
where adaptive synaptic weights [Eq.~\eqref{eqBNcc2}] compensate internal
coalescence and the external input is limited by the finite network size
with excitation probability $\lambda(h)$ for a Poisson input. We thus obtain 
\begin{equation}\label{eqResponseCC}
  a\cc(m,h) = \frac{\lambda(h)}{1-m(1-\lambda(h))},
\end{equation}
for which the inverse is again straightforward
$h\cc(m,a)=-\log\left[1-(1-m)a/(1-ma)\right]/\Delta t$. The dynamic range remains
maximal at criticality (Fig.~\ref{figOverview}~\textbf{C}), while the
discriminable intervals remains depended on $m$
(Fig.~\ref{figOverview}~\textbf{D}). Figure~\ref{figOverview}~\textbf{D}
further shows that our mean-field solution well describes our numerical results
($N=10^2$) and that log-spaced choices of $\epsilon=1-m$ yield a homogeneous
overlap of discriminable intervals.

\begin{figure}
  \includegraphics{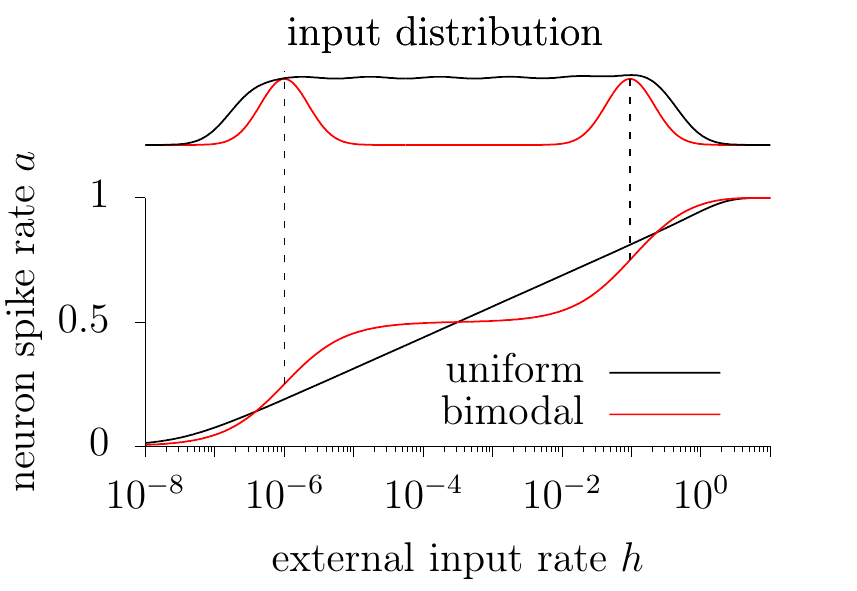}
  \caption{%
    Tailored ensembles of coalescence-compensating networks for particular
    stimulus statistics. Broad uniform distribution of intensities: networks
    with overlapping discriminable intervals. Bimodal distribution of
    intensities: networks with discriminable intervals around the peaks of the
    distribution.
    \label{figEcho}
    }
\end{figure}
%
\first{Our results show that an ensemble of coalescence-compensating networks
can be tailored to optimize sensitivity to stimulus statistics.} For example, a
bimodal distribution of stimulus intensities can be well processed by at least two
networks with disjoint discriminable intervals~(Fig.~\ref{figEcho}), consistent
with the heterogeneous single-neuron responses observed in auditory midbrain when
presented with bimodal sound level intensities~\cite{dean2005}. Processing a
bimodal distribution of stimulus intensities may be also relevant for higher cortical
areas, receiving input from areas with up-and-down states~\cite{wilson2008,
stern1997, holcman2006, millman2010}, or when reacting to complex behavior such
as bimodal escape sequence duration of drosophila~\cite{vonreyn2014}. If,
however, stimulus statistics cannot be anticipated, surprises are best dealt
with by assuming a uniform broad distribution of stimulus intensities. In this case, the best strategy
is to maximize the dynamic range by an ensemble of networks with sufficiently
overlapping discriminable intervals (Fig.~\ref{figOverview}~\textbf{D} and
Fig.~\ref{figEcho}).

\first{The optimized sensitivity of the ensemble to stimulus statistics can be
exploited in machine-learning applications.} A potential application could be a
tailored ensemble of coalescence-compensating recurrent networks in reservoir
computing devices~\cite{buonomano1995, maass2002, jaeger2004, schiller2005,
jaeger2007, boedecker2012}. Stacking these reservoirs will perform optimal
separation of stimulus intensities within the union of discriminable intervals
of participating networks. 

\first{Our results suggest an alternative strategy to solve the discrepancy
between the dynamic range of a single network compared to the large range of
sensory stimulus intensities (dynamic range problem).} First, networks probably
normalize the input statistics, as observed in the early stages of the visual
pathway~\cite{heeger1992}. Second, a single network could implement an adaptable
discriminable interval by tuning its distance to criticality ($m$). Supporting
this possibility, rapid adjustment of neuronal sensitivity to the stimulus
intensity was demonstrated in a recent study of the \emph{Drosophila} hearing
system~\cite{clemens2018}. We hypothesize that the brain combines all strategies
for maximized robustness.


JZ and VP received financial support from the German Ministry of Education and
Research (BMBF) via the Bernstein Center for Computational Neuroscience (BCCN)
G{\"o}ttingen under Grant No.~01GQ1005B. JW was financially supported by
Gertrud-Reemtsma-Stiftung.  AL received funding from a Sofja Kovalevskaja Award
from the Alexander von Humboldt Foundation, endowed by the Federal Ministry of
Education and Research.

\bibliographystyle{./custom.bst} 
\bibliography{./bibliography.bib,addon.bib}

\begin{thebibliography}{41}%
\makeatletter
\providecommand \@ifxundefined [1]{%
 \@ifx{#1\undefined}
}%
\providecommand \@ifnum [1]{%
 \ifnum #1\expandafter \@firstoftwo
 \else \expandafter \@secondoftwo
 \fi
}%
\providecommand \@ifx [1]{%
 \ifx #1\expandafter \@firstoftwo
 \else \expandafter \@secondoftwo
 \fi
}%
\providecommand \natexlab [1]{#1}%
\providecommand \enquote  [1]{``#1''}%
\providecommand \bibnamefont  [1]{#1}%
\providecommand \bibfnamefont [1]{#1}%
\providecommand \citenamefont [1]{#1}%
\providecommand \href@noop [0]{\@secondoftwo}%
\providecommand \href [0]{\begingroup \@sanitize@url \@href}%
\providecommand \@href[1]{\@@startlink{#1}\@@href}%
\providecommand \@@href[1]{\endgroup#1\@@endlink}%
\providecommand \@sanitize@url [0]{\catcode `\\12\catcode `\$12\catcode
  `\&12\catcode `\#12\catcode `\^12\catcode `\_12\catcode `\%12\relax}%
\providecommand \@@startlink[1]{}%
\providecommand \@@endlink[0]{}%
\providecommand \url  [0]{\begingroup\@sanitize@url \@url }%
\providecommand \@url [1]{\endgroup\@href {#1}{\urlprefix }}%
\providecommand \urlprefix  [0]{URL }%
\providecommand \Eprint [0]{\href }%
\providecommand \doibase [0]{http://dx.doi.org/}%
\providecommand \selectlanguage [0]{\@gobble}%
\providecommand \bibinfo  [0]{\@secondoftwo}%
\providecommand \bibfield  [0]{\@secondoftwo}%
\providecommand \translation [1]{[#1]}%
\providecommand \BibitemOpen [0]{}%
\providecommand \bibitemStop [0]{}%
\providecommand \bibitemNoStop [0]{.\EOS\space}%
\providecommand \EOS [0]{\spacefactor3000\relax}%
\providecommand \BibitemShut  [1]{\csname bibitem#1\endcsname}%
\let\auto@bib@innerbib\@empty
\bibitem [{\citenamefont {Hecht}(1924)}]{hecht1924}%
  \BibitemOpen
  \bibfield  {author} {\bibinfo {author} {\bibfnamefont {S.}~\bibnamefont
  {Hecht}},\ }{\selectlanguage {en}\enquote {\bibinfo {title} {The {{Visual
  Discrimination}} of {{Intensity}} and the {{Weber}}-{{Fechner Law}}},}\
  }\href {\doibase 10.1085/jgp.7.2.235} {\bibfield  {journal} {\bibinfo
  {journal} {J. Gen. Physiol.}\ }\textbf {\bibinfo {volume} {7}},\ \bibinfo
  {pages} {235} (\bibinfo {year} {1924})}\BibitemShut {NoStop}%
\bibitem [{\citenamefont {Borg}\ \emph {et~al.}(1967)\citenamefont {Borg},
  \citenamefont {Diamant}, \citenamefont {Str\"om},\ and\ \citenamefont
  {Zotterman}}]{borg1967}%
  \BibitemOpen
  \bibfield  {author} {\bibinfo {author} {\bibfnamefont {G.}~\bibnamefont
  {Borg}}, \bibinfo {author} {\bibfnamefont {H.}~\bibnamefont {Diamant}},
  \bibinfo {author} {\bibfnamefont {L.}~\bibnamefont {Str\"om}}, \ and\
  \bibinfo {author} {\bibfnamefont {Y.}~\bibnamefont {Zotterman}},\
  }{\selectlanguage {en}\enquote {\bibinfo {title} {The relation between neural
  and perceptual intensity: A comparative study on the neural and
  psychophysical response to taste stimuli},}\ }\href {\doibase
  10.1113/jphysiol.1967.sp008284} {\bibfield  {journal} {\bibinfo  {journal}
  {J. Physiol.}\ }\textbf {\bibinfo {volume} {192}},\ \bibinfo {pages} {13}
  (\bibinfo {year} {1967})}\BibitemShut {NoStop}%
\bibitem [{\citenamefont {Viemeister}(1988)}]{viemeister1988}%
  \BibitemOpen
  \bibfield  {author} {\bibinfo {author} {\bibfnamefont {N.~F.}\ \bibnamefont
  {Viemeister}},\ }{\selectlanguage {en}\enquote {\bibinfo {title} {Intensity
  coding and the dynamic range problem},}\ }\href {\doibase
  10.1016/0378-5955(88)90007-X} {\bibfield  {journal} {\bibinfo  {journal}
  {Hear. Res.}\ }\textbf {\bibinfo {volume} {34}},\ \bibinfo {pages} {267}
  (\bibinfo {year} {1988})}\BibitemShut {NoStop}%
\bibitem [{\citenamefont {Rouiller}\ \emph {et~al.}(1983)\citenamefont
  {Rouiller}, \citenamefont {{de Ribaupierre}}, \citenamefont {Morel},\ and\
  \citenamefont {{de Ribaupierre}}}]{rouiller1983}%
  \BibitemOpen
  \bibfield  {author} {\bibinfo {author} {\bibfnamefont {E.}~\bibnamefont
  {Rouiller}}, \bibinfo {author} {\bibfnamefont {Y.}~\bibnamefont {{de
  Ribaupierre}}}, \bibinfo {author} {\bibfnamefont {A.}~\bibnamefont {Morel}},
  \ and\ \bibinfo {author} {\bibfnamefont {F.}~\bibnamefont {{de
  Ribaupierre}}},\ }{\selectlanguage {en}\enquote {\bibinfo {title} {Intensity
  functions of single unit responses to tone in the medial geniculate body of
  cat},}\ }\href {\doibase 10.1016/0378-5955(83)90081-3} {\bibfield  {journal}
  {\bibinfo  {journal} {Hear. Res.}\ }\textbf {\bibinfo {volume} {11}},\
  \bibinfo {pages} {235} (\bibinfo {year} {1983})}\BibitemShut {NoStop}%
\bibitem [{\citenamefont {Viemeister}(1983)}]{viemeister1983}%
  \BibitemOpen
  \bibfield  {author} {\bibinfo {author} {\bibfnamefont {N.}~\bibnamefont
  {Viemeister}},\ }{\selectlanguage {en}\enquote {\bibinfo {title} {Auditory
  intensity discrimination at high frequencies in the presence of noise},}\
  }\href {\doibase 10.1126/science.6612337} {\bibfield  {journal} {\bibinfo
  {journal} {Science}\ }\textbf {\bibinfo {volume} {221}},\ \bibinfo {pages}
  {1206} (\bibinfo {year} {1983})}\BibitemShut {NoStop}%
\bibitem [{\citenamefont {Schreiner}\ \emph {et~al.}(1992)\citenamefont
  {Schreiner}, \citenamefont {Mendelson},\ and\ \citenamefont
  {Sutter}}]{schreiner1992}%
  \BibitemOpen
  \bibfield  {author} {\bibinfo {author} {\bibfnamefont {C.}~\bibnamefont
  {Schreiner}}, \bibinfo {author} {\bibfnamefont {J.}~\bibnamefont
  {Mendelson}}, \ and\ \bibinfo {author} {\bibfnamefont {M.}~\bibnamefont
  {Sutter}},\ }{\selectlanguage {en}\enquote {\bibinfo {title} {Functional
  topography of cat primary auditory cortex: Representation of tone
  intensity},}\ }\href {\doibase 10.1007/BF00230388} {\bibfield  {journal}
  {\bibinfo  {journal} {Exp. Brain Res.}\ }\textbf {\bibinfo {volume} {92}}
  (\bibinfo {year} {1992})}\BibitemShut {NoStop}%
\bibitem [{\citenamefont {Dean}\ \emph {et~al.}(2005)\citenamefont {Dean},
  \citenamefont {Harper},\ and\ \citenamefont {McAlpine}}]{dean2005}%
  \BibitemOpen
  \bibfield  {author} {\bibinfo {author} {\bibfnamefont {I.}~\bibnamefont
  {Dean}}, \bibinfo {author} {\bibfnamefont {N.~S.}\ \bibnamefont {Harper}}, \
  and\ \bibinfo {author} {\bibfnamefont {D.}~\bibnamefont {McAlpine}},\
  }{\selectlanguage {en}\enquote {\bibinfo {title} {Neural population coding of
  sound level adapts to stimulus statistics},}\ }\href {\doibase
  10.1038/nn1541} {\bibfield  {journal} {\bibinfo  {journal} {Nat. Neurosci.}\
  }\textbf {\bibinfo {volume} {8}},\ \bibinfo {pages} {1684} (\bibinfo {year}
  {2005})}\BibitemShut {NoStop}%
\bibitem [{\citenamefont {Evans}\ and\ \citenamefont
  {Palmer}(1980)}]{evans1980}%
  \BibitemOpen
  \bibfield  {author} {\bibinfo {author} {\bibfnamefont {E.}~\bibnamefont
  {Evans}}\ and\ \bibinfo {author} {\bibfnamefont {A.}~\bibnamefont {Palmer}},\
  }{\selectlanguage {en}\enquote {\bibinfo {title} {Relationship between the
  dynamic range of cochlear nerve fibres and their spontaneous activity},}\
  }\href {\doibase 10.1007/BF00236671} {\bibfield  {journal} {\bibinfo
  {journal} {Exp. Brain Res.}\ }\textbf {\bibinfo {volume} {40}} (\bibinfo
  {year} {1980})}\BibitemShut {NoStop}%
\bibitem [{\citenamefont {Evans}(1981)}]{evans1981}%
  \BibitemOpen
  \bibfield  {author} {\bibinfo {author} {\bibfnamefont {E.~F.}\ \bibnamefont
  {Evans}},\ }in\ \href {\doibase 10.1007/978-1-4684-3908-3_9}
  {{\selectlanguage {en}\emph {\bibinfo {booktitle} {Neuronal {{Mechanisms}} of
  {{Hearing}}}}}},\ \bibinfo {editor} {edited by\ \bibinfo {editor}
  {\bibfnamefont {J.}~\bibnamefont {Syka}}\ and\ \bibinfo {editor}
  {\bibfnamefont {L.}~\bibnamefont {Aitkin}}}\ (\bibinfo  {publisher}
  {{Springer US}},\ \bibinfo {address} {Boston, MA},\ \bibinfo {year} {1981})\
  pp.\ \bibinfo {pages} {69--85}\BibitemShut {NoStop}%
\bibitem [{\citenamefont {Kinouchi}\ and\ \citenamefont
  {Copelli}(2006)}]{kinouchi2006}%
  \BibitemOpen
  \bibfield  {author} {\bibinfo {author} {\bibfnamefont {O.}~\bibnamefont
  {Kinouchi}}\ and\ \bibinfo {author} {\bibfnamefont {M.}~\bibnamefont
  {Copelli}},\ }{\selectlanguage {en}\enquote {\bibinfo {title} {Optimal
  dynamical range of excitable networks at criticality},}\ }\href {\doibase
  10.1038/nphys289} {\bibfield  {journal} {\bibinfo  {journal} {Nat. Phys.}\
  }\textbf {\bibinfo {volume} {2}},\ \bibinfo {pages} {348} (\bibinfo {year}
  {2006})}\BibitemShut {NoStop}%
\bibitem [{\citenamefont {Beggs}(2008)}]{beggs2008}%
  \BibitemOpen
  \bibfield  {author} {\bibinfo {author} {\bibfnamefont {J.~M.}\ \bibnamefont
  {Beggs}},\ }{\selectlanguage {en}\enquote {\bibinfo {title} {The criticality
  hypothesis: How local cortical networks might optimize information
  processing},}\ }\href {\doibase 10.1098/rsta.2007.2092} {\bibfield  {journal}
  {\bibinfo  {journal} {Philos. Trans. Royal Soc. A}\ }\textbf {\bibinfo
  {volume} {366}},\ \bibinfo {pages} {329} (\bibinfo {year}
  {2008})}\BibitemShut {NoStop}%
\bibitem [{\citenamefont {Mu\~noz}(2018)}]{munoz2018}%
  \BibitemOpen
  \bibfield  {author} {\bibinfo {author} {\bibfnamefont {M.~A.}\ \bibnamefont
  {Mu\~noz}},\ }\enquote {\bibinfo {title} {Colloquium: {{Criticality}} and
  dynamical scaling in living systems},}\ \href {\doibase
  10.1103/RevModPhys.90.031001} {\bibfield  {journal} {\bibinfo  {journal}
  {Rev. Mod. Phys.}\ }\textbf {\bibinfo {volume} {90}},\ \bibinfo {pages}
  {031001} (\bibinfo {year} {2018})}\BibitemShut {NoStop}%
\bibitem [{\citenamefont {Wu}\ \emph {et~al.}(2007)\citenamefont {Wu},
  \citenamefont {Xu},\ and\ \citenamefont {Wang}}]{wu2007}%
  \BibitemOpen
  \bibfield  {author} {\bibinfo {author} {\bibfnamefont {A.-C.}\ \bibnamefont
  {Wu}}, \bibinfo {author} {\bibfnamefont {X.-J.}\ \bibnamefont {Xu}}, \ and\
  \bibinfo {author} {\bibfnamefont {Y.-H.}\ \bibnamefont {Wang}},\
  }{\selectlanguage {en}\enquote {\bibinfo {title} {Excitable
  {{Greenberg}}-{{Hastings}} cellular automaton model on scale-free
  networks},}\ }\href {\doibase 10.1103/PhysRevE.75.032901} {\bibfield
  {journal} {\bibinfo  {journal} {Phys. Rev. E}\ }\textbf {\bibinfo {volume}
  {75}},\ \bibinfo {pages} {032901} (\bibinfo {year} {2007})}\BibitemShut
  {NoStop}%
\bibitem [{\citenamefont {Larremore}\ \emph {et~al.}(2011)\citenamefont
  {Larremore}, \citenamefont {Shew},\ and\ \citenamefont
  {Restrepo}}]{larremore2011}%
  \BibitemOpen
  \bibfield  {author} {\bibinfo {author} {\bibfnamefont {D.~B.}\ \bibnamefont
  {Larremore}}, \bibinfo {author} {\bibfnamefont {W.~L.}\ \bibnamefont {Shew}},
  \ and\ \bibinfo {author} {\bibfnamefont {J.~G.}\ \bibnamefont {Restrepo}},\
  }{\selectlanguage {en}\enquote {\bibinfo {title} {Predicting {{Criticality}}
  and {{Dynamic Range}} in {{Complex Networks}}: {{Effects}} of
  {{Topology}}},}\ }\href {\doibase 10.1103/PhysRevLett.106.058101} {\bibfield
  {journal} {\bibinfo  {journal} {Phys. Rev. Lett.}\ }\textbf {\bibinfo
  {volume} {106}},\ \bibinfo {pages} {058101} (\bibinfo {year}
  {2011})}\BibitemShut {NoStop}%
\bibitem [{\citenamefont {Shew}\ \emph {et~al.}(2009)\citenamefont {Shew},
  \citenamefont {Yang}, \citenamefont {Petermann}, \citenamefont {Roy},\ and\
  \citenamefont {Plenz}}]{shew2009}%
  \BibitemOpen
  \bibfield  {author} {\bibinfo {author} {\bibfnamefont {W.~L.}\ \bibnamefont
  {Shew}}, \bibinfo {author} {\bibfnamefont {H.}~\bibnamefont {Yang}}, \bibinfo
  {author} {\bibfnamefont {T.}~\bibnamefont {Petermann}}, \bibinfo {author}
  {\bibfnamefont {R.}~\bibnamefont {Roy}}, \ and\ \bibinfo {author}
  {\bibfnamefont {D.}~\bibnamefont {Plenz}},\ }{\selectlanguage {en}\enquote
  {\bibinfo {title} {Neuronal {{Avalanches Imply Maximum Dynamic Range}} in
  {{Cortical Networks}} at {{Criticality}}},}\ }\href {\doibase
  10.1523/JNEUROSCI.3864-09.2009} {\bibfield  {journal} {\bibinfo  {journal}
  {J. Neurosci.}\ }\textbf {\bibinfo {volume} {29}},\ \bibinfo {pages} {15595}
  (\bibinfo {year} {2009})}\BibitemShut {NoStop}%
\bibitem [{\citenamefont {Chevallier}(2018)}]{chevallier2018}%
  \BibitemOpen
  \bibfield  {author} {\bibinfo {author} {\bibfnamefont {J.}~\bibnamefont
  {Chevallier}},\ }{\selectlanguage {en}\enquote {\bibinfo {title} {Stimulus
  {{Sensitivity}} of a {{Spiking Neural Network Model}}},}\ }\href {\doibase
  10.1007/s10955-017-1948-y} {\bibfield  {journal} {\bibinfo  {journal} {J.
  Stat. Phys.}\ }\textbf {\bibinfo {volume} {170}},\ \bibinfo {pages} {800}
  (\bibinfo {year} {2018})}\BibitemShut {NoStop}%
\bibitem [{\citenamefont {Haldeman}\ and\ \citenamefont
  {Beggs}(2005)}]{haldeman2005}%
  \BibitemOpen
  \bibfield  {author} {\bibinfo {author} {\bibfnamefont {C.}~\bibnamefont
  {Haldeman}}\ and\ \bibinfo {author} {\bibfnamefont {J.~M.}\ \bibnamefont
  {Beggs}},\ }{\selectlanguage {en}\enquote {\bibinfo {title} {Critical
  {{Branching Captures Activity}} in {{Living Neural Networks}} and
  {{Maximizes}} the {{Number}} of {{Metastable States}}},}\ }\href {\doibase
  10.1103/PhysRevLett.94.058101} {\bibfield  {journal} {\bibinfo  {journal}
  {Phys. Rev. Lett.}\ }\textbf {\bibinfo {volume} {94}} (\bibinfo {year}
  {2005})}\BibitemShut {NoStop}%
\bibitem [{\citenamefont {Zierenberg}\ \emph
  {et~al.}(2018{\natexlab{a}})\citenamefont {Zierenberg}, \citenamefont
  {Wilting},\ and\ \citenamefont {Priesemann}}]{zierenberg2018}%
  \BibitemOpen
  \bibfield  {author} {\bibinfo {author} {\bibfnamefont {J.}~\bibnamefont
  {Zierenberg}}, \bibinfo {author} {\bibfnamefont {J.}~\bibnamefont {Wilting}},
  \ and\ \bibinfo {author} {\bibfnamefont {V.}~\bibnamefont {Priesemann}},\
  }{\selectlanguage {en}\enquote {\bibinfo {title} {Homeostatic {{Plasticity}}
  and {{External Input Shape Neural Network Dynamics}}},}\ }\href {\doibase
  10.1103/PhysRevX.8.031018} {\bibfield  {journal} {\bibinfo  {journal} {Phys.
  Rev. X}\ }\textbf {\bibinfo {volume} {8}} (\bibinfo {year}
  {2018}{\natexlab{a}})}\BibitemShut {NoStop}%
\bibitem [{\citenamefont {Zierenberg}\ \emph
  {et~al.}(2018{\natexlab{b}})\citenamefont {Zierenberg}, \citenamefont
  {Wilting}, \citenamefont {Priesemann},\ and\ \citenamefont
  {Levina}}]{zierenbergLong}%
  \BibitemOpen
  \bibfield  {author} {\bibinfo {author} {\bibfnamefont {J.}~\bibnamefont
  {Zierenberg}}, \bibinfo {author} {\bibfnamefont {J.}~\bibnamefont {Wilting}},
  \bibinfo {author} {\bibfnamefont {V.}~\bibnamefont {Priesemann}}, \ and\
  \bibinfo {author} {\bibfnamefont {A.}~\bibnamefont {Levina}},\ }\enquote
  {\bibinfo {title} {Description of spreading dynamics by microscopic network
  models and macroscopic branching processes can differ due to coalescence},}\
  \href@noop {} {\bibfield  {journal} {\bibinfo  {journal} {submitted}\ }
  (\bibinfo {year} {2018}{\natexlab{b}})}\BibitemShut {NoStop}%
\bibitem [{\citenamefont {Corless}\ \emph {et~al.}(1996)\citenamefont
  {Corless}, \citenamefont {Gonnet}, \citenamefont {Hare}, \citenamefont
  {Jeffrey},\ and\ \citenamefont {Knuth}}]{corless1996}%
  \BibitemOpen
  \bibfield  {author} {\bibinfo {author} {\bibfnamefont {R.~M.}\ \bibnamefont
  {Corless}}, \bibinfo {author} {\bibfnamefont {G.~H.}\ \bibnamefont {Gonnet}},
  \bibinfo {author} {\bibfnamefont {D.~E.~G.}\ \bibnamefont {Hare}}, \bibinfo
  {author} {\bibfnamefont {D.~J.}\ \bibnamefont {Jeffrey}}, \ and\ \bibinfo
  {author} {\bibfnamefont {D.~E.}\ \bibnamefont {Knuth}},\ }{\selectlanguage
  {en}\enquote {\bibinfo {title} {On the {{LambertW}} function},}\ }\href
  {\doibase 10.1007/BF02124750} {\bibfield  {journal} {\bibinfo  {journal}
  {Adv. Comput. Math.}\ }\textbf {\bibinfo {volume} {5}},\ \bibinfo {pages}
  {329} (\bibinfo {year} {1996})}\BibitemShut {NoStop}%
\bibitem [{\citenamefont {Wilting}\ and\ \citenamefont
  {Priesemann}(2018)}]{wilting2018}%
  \BibitemOpen
  \bibfield  {author} {\bibinfo {author} {\bibfnamefont {J.}~\bibnamefont
  {Wilting}}\ and\ \bibinfo {author} {\bibfnamefont {V.}~\bibnamefont
  {Priesemann}},\ }{\selectlanguage {en}\enquote {\bibinfo {title} {Inferring
  collective dynamical states from widely unobserved systems},}\ }\href
  {\doibase 10.1038/s41467-018-04725-4} {\bibfield  {journal} {\bibinfo
  {journal} {Nat. Commun.}\ }\textbf {\bibinfo {volume} {9}},\ \bibinfo {pages}
  {2325} (\bibinfo {year} {2018})}\BibitemShut {NoStop}%
\bibitem [{\citenamefont {Wilting}\ \emph {et~al.}(2018)\citenamefont
  {Wilting}, \citenamefont {Dehning}, \citenamefont {Pinheiro~Neto},
  \citenamefont {Rudelt}, \citenamefont {Wibral}, \citenamefont {Zierenberg},\
  and\ \citenamefont {Priesemann}}]{wilting2018a}%
  \BibitemOpen
  \bibfield  {author} {\bibinfo {author} {\bibfnamefont {J.}~\bibnamefont
  {Wilting}}, \bibinfo {author} {\bibfnamefont {J.}~\bibnamefont {Dehning}},
  \bibinfo {author} {\bibfnamefont {J.}~\bibnamefont {Pinheiro~Neto}}, \bibinfo
  {author} {\bibfnamefont {L.}~\bibnamefont {Rudelt}}, \bibinfo {author}
  {\bibfnamefont {M.}~\bibnamefont {Wibral}}, \bibinfo {author} {\bibfnamefont
  {J.}~\bibnamefont {Zierenberg}}, \ and\ \bibinfo {author} {\bibfnamefont
  {V.}~\bibnamefont {Priesemann}},\ }{\selectlanguage {English}\enquote
  {\bibinfo {title} {Operating in a {{Reverberating Regime Enables Rapid
  Tuning}} of {{Network States}} to {{Task Requirements}}},}\ }\href {\doibase
  10.3389/fnsys.2018.00055} {\bibfield  {journal} {\bibinfo  {journal} {Front.
  Syst. Neurosci.}\ }\textbf {\bibinfo {volume} {12}} (\bibinfo {year}
  {2018})}\BibitemShut {NoStop}%
\bibitem [{\citenamefont {Harris}(1963)}]{harris1963}%
  \BibitemOpen
  \bibfield  {author} {\bibinfo {author} {\bibfnamefont {T.~E.}\ \bibnamefont
  {Harris}},\ }\href@noop {} {{\selectlanguage {en}\emph {\bibinfo {title} {The
  {{Theory}} of {{Branching Processes}}}}}}\ (\bibinfo  {publisher} {{Springer
  Belrin}},\ \bibinfo {year} {1963})\BibitemShut {NoStop}%
\bibitem [{\citenamefont {Beggs}\ and\ \citenamefont
  {Plenz}(2003)}]{beggs2003}%
  \BibitemOpen
  \bibfield  {author} {\bibinfo {author} {\bibfnamefont {J.~M.}\ \bibnamefont
  {Beggs}}\ and\ \bibinfo {author} {\bibfnamefont {D.}~\bibnamefont {Plenz}},\
  }{\selectlanguage {en}\enquote {\bibinfo {title} {Neuronal {{Avalanches}} in
  {{Neocortical Circuits}}},}\ }\href {\doibase
  10.1523/JNEUROSCI.23-35-11167.2003} {\bibfield  {journal} {\bibinfo
  {journal} {J. Neurosci.}\ }\textbf {\bibinfo {volume} {23}},\ \bibinfo
  {pages} {11167} (\bibinfo {year} {2003})}\BibitemShut {NoStop}%
\bibitem [{\citenamefont {Priesemann}\ \emph {et~al.}(2014)\citenamefont
  {Priesemann}, \citenamefont {Wibral}, \citenamefont {Valderrama},
  \citenamefont {Pr\"opper}, \citenamefont {Le~Van~Quyen}, \citenamefont
  {Geisel}, \citenamefont {Triesch}, \citenamefont {Nikoli\'c},\ and\
  \citenamefont {Munk}}]{priesemann2014}%
  \BibitemOpen
  \bibfield  {author} {\bibinfo {author} {\bibfnamefont {V.}~\bibnamefont
  {Priesemann}}, \bibinfo {author} {\bibfnamefont {M.}~\bibnamefont {Wibral}},
  \bibinfo {author} {\bibfnamefont {M.}~\bibnamefont {Valderrama}}, \bibinfo
  {author} {\bibfnamefont {R.}~\bibnamefont {Pr\"opper}}, \bibinfo {author}
  {\bibfnamefont {M.}~\bibnamefont {Le~Van~Quyen}}, \bibinfo {author}
  {\bibfnamefont {T.}~\bibnamefont {Geisel}}, \bibinfo {author} {\bibfnamefont
  {J.}~\bibnamefont {Triesch}}, \bibinfo {author} {\bibfnamefont
  {D.}~\bibnamefont {Nikoli\'c}}, \ and\ \bibinfo {author} {\bibfnamefont
  {M.~H.~J.}\ \bibnamefont {Munk}},\ }{\selectlanguage {English}\enquote
  {\bibinfo {title} {Spike avalanches in vivo suggest a driven, slightly
  subcritical brain state},}\ }\href {\doibase 10.3389/fnsys.2014.00108}
  {\bibfield  {journal} {\bibinfo  {journal} {Front. Syst. Neurosci.}\ }\textbf
  {\bibinfo {volume} {8}} (\bibinfo {year} {2014})}\BibitemShut {NoStop}%
\bibitem [{\citenamefont {Levina}\ \emph {et~al.}(2008)\citenamefont {Levina},
  \citenamefont {Herrmann},\ and\ \citenamefont {Denker}}]{levina2008}%
  \BibitemOpen
  \bibfield  {author} {\bibinfo {author} {\bibfnamefont {A.}~\bibnamefont
  {Levina}}, \bibinfo {author} {\bibfnamefont {J.~M.}\ \bibnamefont
  {Herrmann}}, \ and\ \bibinfo {author} {\bibfnamefont {M.}~\bibnamefont
  {Denker}},\ }{\selectlanguage {en}\enquote {\bibinfo {title} {Critical
  branching processes in neural networks},}\ }\href {\doibase
  10.1002/pamm.200700029} {\bibfield  {journal} {\bibinfo  {journal} {PAMM}\
  }\textbf {\bibinfo {volume} {7}},\ \bibinfo {pages} {1030701} (\bibinfo
  {year} {2008})}\BibitemShut {NoStop}%
\bibitem [{\citenamefont {Levina}\ \emph {et~al.}(2007)\citenamefont {Levina},
  \citenamefont {Ernst},\ and\ \citenamefont {Michael~Herrmann}}]{levina2007}%
  \BibitemOpen
  \bibfield  {author} {\bibinfo {author} {\bibfnamefont {A.}~\bibnamefont
  {Levina}}, \bibinfo {author} {\bibfnamefont {U.}~\bibnamefont {Ernst}}, \
  and\ \bibinfo {author} {\bibfnamefont {J.}~\bibnamefont {Michael~Herrmann}},\
  }{\selectlanguage {en}\enquote {\bibinfo {title} {Criticality of avalanche
  dynamics in adaptive recurrent networks},}\ }\href {\doibase
  10.1016/j.neucom.2006.10.056} {\bibfield  {journal} {\bibinfo  {journal}
  {Neurocomputing}\ }\textbf {\bibinfo {volume} {70}},\ \bibinfo {pages} {1877}
  (\bibinfo {year} {2007})}\BibitemShut {NoStop}%
\bibitem [{\citenamefont {Friedman}\ \emph {et~al.}(2012)\citenamefont
  {Friedman}, \citenamefont {Ito}, \citenamefont {Brinkman}, \citenamefont
  {Shimono}, \citenamefont {DeVille}, \citenamefont {Dahmen}, \citenamefont
  {Beggs},\ and\ \citenamefont {Butler}}]{friedman2012}%
  \BibitemOpen
  \bibfield  {author} {\bibinfo {author} {\bibfnamefont {N.}~\bibnamefont
  {Friedman}}, \bibinfo {author} {\bibfnamefont {S.}~\bibnamefont {Ito}},
  \bibinfo {author} {\bibfnamefont {B.~A.~W.}\ \bibnamefont {Brinkman}},
  \bibinfo {author} {\bibfnamefont {M.}~\bibnamefont {Shimono}}, \bibinfo
  {author} {\bibfnamefont {R.~E.~L.}\ \bibnamefont {DeVille}}, \bibinfo
  {author} {\bibfnamefont {K.~A.}\ \bibnamefont {Dahmen}}, \bibinfo {author}
  {\bibfnamefont {J.~M.}\ \bibnamefont {Beggs}}, \ and\ \bibinfo {author}
  {\bibfnamefont {T.~C.}\ \bibnamefont {Butler}},\ }{\selectlanguage
  {en}\enquote {\bibinfo {title} {Universal {{Critical Dynamics}} in {{High
  Resolution Neuronal Avalanche Data}}},}\ }\href {\doibase
  10.1103/PhysRevLett.108.208102} {\bibfield  {journal} {\bibinfo  {journal}
  {Phys. Rev. Lett.}\ }\textbf {\bibinfo {volume} {108}},\ \bibinfo {pages}
  {208102} (\bibinfo {year} {2012})}\BibitemShut {NoStop}%
\bibitem [{\citenamefont {Wilson}(2008)}]{wilson2008}%
  \BibitemOpen
  \bibfield  {author} {\bibinfo {author} {\bibfnamefont {C.}~\bibnamefont
  {Wilson}},\ }{\selectlanguage {en}\enquote {\bibinfo {title} {Up and down
  states},}\ }\href@noop {} {\bibfield  {journal} {\bibinfo  {journal}
  {Scholarpedia J}\ }\textbf {\bibinfo {volume} {3}},\ \bibinfo {pages} {1410}
  (\bibinfo {year} {2008})}\BibitemShut {NoStop}%
\bibitem [{\citenamefont {Stern}\ \emph {et~al.}(1997)\citenamefont {Stern},
  \citenamefont {Kincaid},\ and\ \citenamefont {Wilson}}]{stern1997}%
  \BibitemOpen
  \bibfield  {author} {\bibinfo {author} {\bibfnamefont {E.~A.}\ \bibnamefont
  {Stern}}, \bibinfo {author} {\bibfnamefont {A.~E.}\ \bibnamefont {Kincaid}},
  \ and\ \bibinfo {author} {\bibfnamefont {C.~J.}\ \bibnamefont {Wilson}},\
  }{\selectlanguage {en}\enquote {\bibinfo {title} {Spontaneous {{Subthreshold
  Membrane Potential Fluctuations}} and {{Action Potential Variability}} of
  {{Rat Corticostriatal}} and {{Striatal Neurons In Vivo}}},}\ }\href {\doibase
  10.1152/jn.1997.77.4.1697} {\bibfield  {journal} {\bibinfo  {journal} {J.
  Neurophysiol.}\ }\textbf {\bibinfo {volume} {77}},\ \bibinfo {pages} {1697}
  (\bibinfo {year} {1997})}\BibitemShut {NoStop}%
\bibitem [{\citenamefont {Holcman}\ and\ \citenamefont
  {Tsodyks}(2006)}]{holcman2006}%
  \BibitemOpen
  \bibfield  {author} {\bibinfo {author} {\bibfnamefont {D.}~\bibnamefont
  {Holcman}}\ and\ \bibinfo {author} {\bibfnamefont {M.}~\bibnamefont
  {Tsodyks}},\ }{\selectlanguage {en}\enquote {\bibinfo {title} {The
  {{Emergence}} of {{Up}} and {{Down States}} in {{Cortical Networks}}},}\
  }\href {\doibase 10.1371/journal.pcbi.0020023} {\bibfield  {journal}
  {\bibinfo  {journal} {PLoS Comput. Biol.}\ }\textbf {\bibinfo {volume} {2}},\
  \bibinfo {pages} {e23} (\bibinfo {year} {2006})}\BibitemShut {NoStop}%
\bibitem [{\citenamefont {Millman}\ \emph {et~al.}(2010)\citenamefont
  {Millman}, \citenamefont {Mihalas}, \citenamefont {Kirkwood},\ and\
  \citenamefont {Niebur}}]{millman2010}%
  \BibitemOpen
  \bibfield  {author} {\bibinfo {author} {\bibfnamefont {D.}~\bibnamefont
  {Millman}}, \bibinfo {author} {\bibfnamefont {S.}~\bibnamefont {Mihalas}},
  \bibinfo {author} {\bibfnamefont {A.}~\bibnamefont {Kirkwood}}, \ and\
  \bibinfo {author} {\bibfnamefont {E.}~\bibnamefont {Niebur}},\
  }{\selectlanguage {en}\enquote {\bibinfo {title} {Self-organized criticality
  occurs in non-conservative neuronal networks during `up' states},}\ }\href
  {\doibase 10.1038/nphys1757} {\bibfield  {journal} {\bibinfo  {journal} {Nat.
  Phys.}\ }\textbf {\bibinfo {volume} {6}},\ \bibinfo {pages} {801} (\bibinfo
  {year} {2010})}\BibitemShut {NoStop}%
\bibitem [{\citenamefont {{von Reyn}}\ \emph {et~al.}(2014)\citenamefont {{von
  Reyn}}, \citenamefont {Breads}, \citenamefont {Peek}, \citenamefont {Zheng},
  \citenamefont {Williamson}, \citenamefont {Yee}, \citenamefont {Leonardo},\
  and\ \citenamefont {Card}}]{vonreyn2014}%
  \BibitemOpen
  \bibfield  {author} {\bibinfo {author} {\bibfnamefont {C.~R.}\ \bibnamefont
  {{von Reyn}}}, \bibinfo {author} {\bibfnamefont {P.}~\bibnamefont {Breads}},
  \bibinfo {author} {\bibfnamefont {M.~Y.}\ \bibnamefont {Peek}}, \bibinfo
  {author} {\bibfnamefont {G.~Z.}\ \bibnamefont {Zheng}}, \bibinfo {author}
  {\bibfnamefont {W.~R.}\ \bibnamefont {Williamson}}, \bibinfo {author}
  {\bibfnamefont {A.~L.}\ \bibnamefont {Yee}}, \bibinfo {author} {\bibfnamefont
  {A.}~\bibnamefont {Leonardo}}, \ and\ \bibinfo {author} {\bibfnamefont
  {G.~M.}\ \bibnamefont {Card}},\ }{\selectlanguage {en}\enquote {\bibinfo
  {title} {A spike-timing mechanism for action selection},}\ }\href {\doibase
  10.1038/nn.3741} {\bibfield  {journal} {\bibinfo  {journal} {Nat. Neurosci.}\
  }\textbf {\bibinfo {volume} {17}},\ \bibinfo {pages} {962} (\bibinfo {year}
  {2014})}\BibitemShut {NoStop}%
\bibitem [{\citenamefont {Buonomano}\ and\ \citenamefont
  {Merzenich}(1995)}]{buonomano1995}%
  \BibitemOpen
  \bibfield  {author} {\bibinfo {author} {\bibfnamefont {D.~V.}\ \bibnamefont
  {Buonomano}}\ and\ \bibinfo {author} {\bibfnamefont {M.~M.}\ \bibnamefont
  {Merzenich}},\ }{\selectlanguage {en}\enquote {\bibinfo {title} {Temporal
  information transformed into a spatial code by a neural network with
  realistic properties},}\ }\href {\doibase 10.1126/science.7863330} {\bibfield
   {journal} {\bibinfo  {journal} {Science}\ }\textbf {\bibinfo {volume}
  {267}},\ \bibinfo {pages} {1028} (\bibinfo {year} {1995})}\BibitemShut
  {NoStop}%
\bibitem [{\citenamefont {Maass}\ \emph {et~al.}(2002)\citenamefont {Maass},
  \citenamefont {Natschl\"ager},\ and\ \citenamefont {Markram}}]{maass2002}%
  \BibitemOpen
  \bibfield  {author} {\bibinfo {author} {\bibfnamefont {W.}~\bibnamefont
  {Maass}}, \bibinfo {author} {\bibfnamefont {T.}~\bibnamefont
  {Natschl\"ager}}, \ and\ \bibinfo {author} {\bibfnamefont {H.}~\bibnamefont
  {Markram}},\ }{\selectlanguage {en}\enquote {\bibinfo {title} {Real-{{Time
  Computing Without Stable States}}: {{A New Framework}} for {{Neural
  Computation Based}} on {{Perturbations}}},}\ }\href {\doibase
  10.1162/089976602760407955} {\bibfield  {journal} {\bibinfo  {journal}
  {Neural Comput.}\ }\textbf {\bibinfo {volume} {14}},\ \bibinfo {pages} {2531}
  (\bibinfo {year} {2002})}\BibitemShut {NoStop}%
\bibitem [{\citenamefont {Jaeger}\ and\ \citenamefont
  {Haas}(2004)}]{jaeger2004}%
  \BibitemOpen
  \bibfield  {author} {\bibinfo {author} {\bibfnamefont {H.}~\bibnamefont
  {Jaeger}}\ and\ \bibinfo {author} {\bibfnamefont {H.}~\bibnamefont {Haas}},\
  }{\selectlanguage {en}\enquote {\bibinfo {title} {Harnessing
  {{Nonlinearity}}: {{Predicting Chaotic Systems}} and {{Saving Energy}} in
  {{Wireless Communication}}},}\ }\href {\doibase 10.1126/science.1091277}
  {\bibfield  {journal} {\bibinfo  {journal} {Science}\ }\textbf {\bibinfo
  {volume} {304}},\ \bibinfo {pages} {78} (\bibinfo {year} {2004})}\BibitemShut
  {NoStop}%
\bibitem [{\citenamefont {Schiller}\ and\ \citenamefont
  {Steil}(2005)}]{schiller2005}%
  \BibitemOpen
  \bibfield  {author} {\bibinfo {author} {\bibfnamefont {U.~D.}\ \bibnamefont
  {Schiller}}\ and\ \bibinfo {author} {\bibfnamefont {J.~J.}\ \bibnamefont
  {Steil}},\ }{\selectlanguage {en}\enquote {\bibinfo {title} {Analyzing the
  weight dynamics of recurrent learning algorithms},}\ }\href {\doibase
  10.1016/j.neucom.2004.04.006} {\bibfield  {journal} {\bibinfo  {journal}
  {Neurocomputing}\ }\textbf {\bibinfo {volume} {63}},\ \bibinfo {pages} {5}
  (\bibinfo {year} {2005})}\BibitemShut {NoStop}%
\bibitem [{\citenamefont {Jaeger}\ \emph {et~al.}(2007)\citenamefont {Jaeger},
  \citenamefont {Maass},\ and\ \citenamefont {Principe}}]{jaeger2007}%
  \BibitemOpen
  \bibfield  {author} {\bibinfo {author} {\bibfnamefont {H.}~\bibnamefont
  {Jaeger}}, \bibinfo {author} {\bibfnamefont {W.}~\bibnamefont {Maass}}, \
  and\ \bibinfo {author} {\bibfnamefont {J.}~\bibnamefont {Principe}},\
  }{\selectlanguage {en}\enquote {\bibinfo {title} {Special issue on echo state
  networks and liquid state machines},}\ }\href {\doibase
  10.1016/j.neunet.2007.04.001} {\bibfield  {journal} {\bibinfo  {journal}
  {Neural Netw.}\ }\textbf {\bibinfo {volume} {20}},\ \bibinfo {pages} {287}
  (\bibinfo {year} {2007})}\BibitemShut {NoStop}%
\bibitem [{\citenamefont {Boedecker}\ \emph {et~al.}(2012)\citenamefont
  {Boedecker}, \citenamefont {Obst}, \citenamefont {Lizier}, \citenamefont
  {Mayer},\ and\ \citenamefont {Asada}}]{boedecker2012}%
  \BibitemOpen
  \bibfield  {author} {\bibinfo {author} {\bibfnamefont {J.}~\bibnamefont
  {Boedecker}}, \bibinfo {author} {\bibfnamefont {O.}~\bibnamefont {Obst}},
  \bibinfo {author} {\bibfnamefont {J.~T.}\ \bibnamefont {Lizier}}, \bibinfo
  {author} {\bibfnamefont {N.~M.}\ \bibnamefont {Mayer}}, \ and\ \bibinfo
  {author} {\bibfnamefont {M.}~\bibnamefont {Asada}},\ }{\selectlanguage
  {en}\enquote {\bibinfo {title} {Information processing in echo state networks
  at the edge of chaos},}\ }\href {\doibase 10.1007/s12064-011-0146-8}
  {\bibfield  {journal} {\bibinfo  {journal} {Theory Biosci.}\ }\textbf
  {\bibinfo {volume} {131}},\ \bibinfo {pages} {205} (\bibinfo {year}
  {2012})}\BibitemShut {NoStop}%
\bibitem [{\citenamefont {Heeger}(1992)}]{heeger1992}%
  \BibitemOpen
  \bibfield  {author} {\bibinfo {author} {\bibfnamefont {D.~J.}\ \bibnamefont
  {Heeger}},\ }{\selectlanguage {en}\enquote {\bibinfo {title} {Normalization
  of cell responses in cat striate cortex},}\ }\href {\doibase
  10.1017/S0952523800009640} {\bibfield  {journal} {\bibinfo  {journal} {Vis.
  Neurosci.}\ }\textbf {\bibinfo {volume} {9}},\ \bibinfo {pages} {181}
  (\bibinfo {year} {1992})}\BibitemShut {NoStop}%
\bibitem [{\citenamefont {Clemens}\ \emph {et~al.}(2018)\citenamefont
  {Clemens}, \citenamefont {{Ozeri-Engelhard}},\ and\ \citenamefont
  {Murthy}}]{clemens2018}%
  \BibitemOpen
  \bibfield  {author} {\bibinfo {author} {\bibfnamefont {J.}~\bibnamefont
  {Clemens}}, \bibinfo {author} {\bibfnamefont {N.}~\bibnamefont
  {{Ozeri-Engelhard}}}, \ and\ \bibinfo {author} {\bibfnamefont
  {M.}~\bibnamefont {Murthy}},\ }{\selectlanguage {en}\enquote {\bibinfo
  {title} {Fast intensity adaptation enhances the encoding of sound in
  {{Drosophila}}},}\ }\href {\doibase 10.1038/s41467-017-02453-9} {\bibfield
  {journal} {\bibinfo  {journal} {Nat. Commun.}\ }\textbf {\bibinfo {volume}
  {9}},\ \bibinfo {pages} {134} (\bibinfo {year} {2018})}\BibitemShut {NoStop}%
\end{thebibliography}%

\onecolumngrid
\clearpage
\begin{center}
\textbf{\large Supplemental Material:\\ Tailored ensembles of neural networks optimize sensitivity to stimulus statistics}
\end{center}

\begin{center}
  J. Zierenberg$^{1,2}$, J. Wilting$^{1}$, V. Priesemann$^{1,2}$, and A. Levina$^{3,4}$\\
{\em \small 
  $^1$ Max Planck Institute for Dynamics and Self-Organization, Am Fassberg 17, 37077 G{\"o}ttingen, Germany,\\
  $^2$ Bernstein Center for Computational Neuroscience, Am Fassberg 17, 37077 G{\"o}ttingen, Germany,\\
  $^3$ University of T\"ubingen, Max Planck Ring 8, 72076 T\"ubingen, Germany,\\
  $^4$ \mbox{Max Planck Institute for Biological Cybernetics, Max Planck Ring 8, 72076 T\"ubingen, Germany}
  }
\end{center}
\twocolumngrid
\setcounter{equation}{0}
\setcounter{figure}{0}
\setcounter{table}{0}
\setcounter{page}{1}
\setcounter{section}{1}
\setcounter{subsection}{1}
\makeatletter
\renewcommand{\theequation}{S\arabic{equation}}
\renewcommand{\thefigure}{S\arabic{figure}}
\renewcommand{\thesubsection}{S.\Roman{subsection}}
\subsection{S1. Adaptive weights compensate internal coalescence}
\label{secResultCC}
Internal and external coalescence reduce the effective branching parameter for
static connection weights $w=m/N$~\cite{zierenbergLong}. Thereby, macroscopic
branching parameters $\hat{m}$, estimated from the network rate, differ from the
model branching parameter $m$. For a detailed discussion, we refer to
Ref.~\cite{zierenbergLong}. We will now exploit the analytical insight on
coalescence to construct microscopic dynamics that compensate for internal
coalescence. The basic idea is simple: adjust the microscopic dynamics ($w$)
such that the effective branching parameter matches the desired macroscopic
branching parameter.

To compensate for coalescence, we adjust the weights $w$ such that
$\meff(w, A_t)=\mtarget$ for all $A_t$, and thereby tune the model parameter
equal to the macroscopic branching parameter $\mtarget=\hat{m}$. Inserting
Eq.~\eqref{eqBNconvergence}, we obtain activity-dependent weights
\begin{equation}
  \widetilde{w}_\mathrm{cc}(A_t) = 1-\left(1-\frac{\mtarget A_t}{N(1-\lambda(h))}\right)^{1/A_t},
\end{equation}
which would compensate for internal and external coalescence. We now assume that
the network has a mechanism to communicate the current activity $A_t$ to each
neuron, but that it cannot have information about the external input rate $h$.  As
a result, we neglect the factor $(1-\lambda(h))$ and obtain the
adaptive (coalescence-compensating) weights 
\begin{equation}\label{eqBNfsc}
  \wcc(m,A_t) = 1-\left(1-\frac{\mtarget A_t}{N}\right)^{1/A_t},
\end{equation}
which compensate only for internal coalescence. Inserted as weight in
Eq.~\eqref{eqBNconvergence}, the remaining external coalescence leads to
$\meff(\wcc(m,A_t),A_t) = \mtarget(1-\lambda(h))$, cf.~Eq.~\eqref{eqBNcc2},
which determines the $N\to\infty$ limit of linear-regression
estimates~\cite{zierenbergLong}. 

\begin{figure}[t]
  \includegraphics{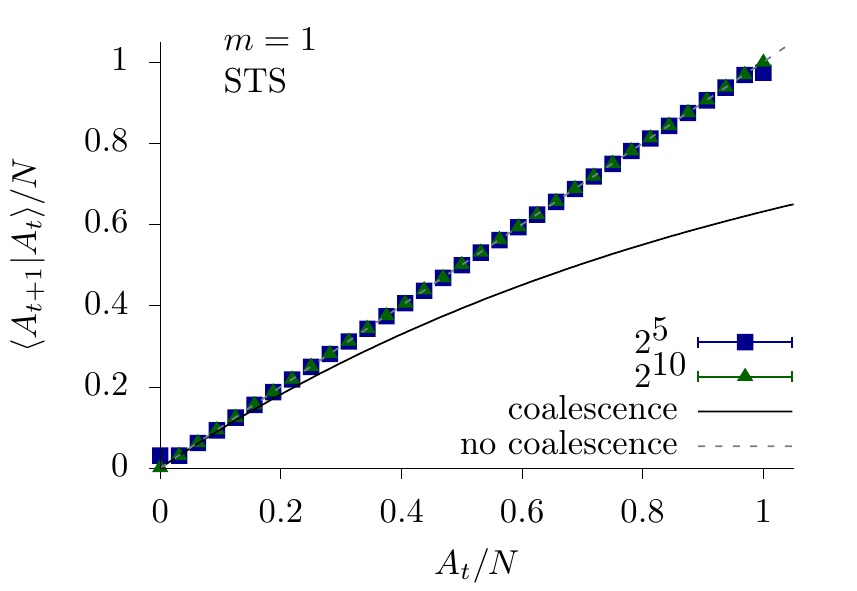}
  \caption{%
    Universal scaling function for effective spreading of network activity as
    discussed in Ref.~\cite{zierenbergLong} ($m=1$ in the
    separation-of-timescale regime). For the branching network (solid line),
    the scaling function is non-linear, causing a bias in linear estimates of
    the branching parameter through network rate. However, for
    coalescence-compensating networks (data points), the conditional
    expectation value is again linear and is described by the scaling function
    of a branching process (dashed line). 
    \label{figResultScalingFunctionCC}
    }
\end{figure}

The adaptive (activity-dependent) weight, Eq.~\eqref{eqBNfsc}, bridges the gap
to the macroscopic description (Fig.~\ref{figResultScalingFunctionCC}). The conditional expectation
value for the coalescence-compensating network with adaptive weights now coincides
with that of a branching process (dashed line), different from the conditional
expectation value of the branching network subject to coalescence (solid line),
see Ref.~\cite{zierenbergLong}. 


We note that Eq.~\eqref{eqBNfsc} is well defined only for $\mtarget<1$, because
for $\mtarget=1$ the adaptive weights diverge at $A_t=N$. This is explained by
the fact that for critical-like dynamics ($\mtarget=1$) the compensation of
coalescence induces an avalanche-size distribution with a perfect power law
behavior. A perfect power law, however, includes avalanches of all sizes. For a
finite network with coalescence compensation this mathematically results in a
second absorbing state at full network activity. 

We need to avoid the absorbing state at full network activity for
$\mtarget\to1$, because the microscopic dynamics would not allow to ever leave
this state again.  For this, we will derive a \textit{cutoff weight}
$\wcc(m,N)=\wcc^\dagger$ such that the probability to transition away from
$A_t=N$ is non-zero. 

To derive the cutoff weight, we need to calculate the probability that all
neurons are activated, if in the previous time step already all neurons were
active. Recall $P\left[s^i_t=1|
A_t,w\cc,h\right]=1-\left(1-w\cc\right)^{A_t}\left(1-\lambda(h)\right)$ for the
probability to activate a single neuron given that $A_t$ neurons are active.
Now, we restrict our discussion to the divergent case $mA_t=N$ in
Eq.~\eqref{eqBNfsc} (defining for $m=1$ $w\cc^\dagger=w\cc(1,N)$) and further neglect the external
input rate ($h\to 0$), because the divergence only occurs for $\mtarget\to 1$,
where the external input rate has to be small for persistent activity. Then, the
probability to activate all neurons in the network, given that all neurons where
activated already in the previous time step is 
\begin{equation}\label{eqPall}
  p_\mathrm{all}=\left(1-\left(1-\wcc^\dagger\right)^N\right)^N.
\end{equation}
The probability to transition away from $A_t=N$ is given by the probability to
not activate at least one neuron, i.e., by $1-p_\mathrm{all}$.

We now ask, how to choose $\wcc^\dagger$ such that once the network is fully active,
there is a non-vanishing probability to transition away from $A_t=N$, i.e.,
that $p_\mathrm{all}<1$. For this we solve Eq.~\eqref{eqPall} for $\wcc^\dagger$ and
obtain 
\begin{align}\label{eqAnsatzCutoff}
  \wcc^\dagger 
  &= 1-\left(1-p_\mathrm{all}^{1/N}\right)^{1/N}\nonumber\\
  &= 1-\left(1-e^{\alpha/N}\right)^{1/N},
\end{align}
where we have introduced $\alpha=\ln(p_\mathrm{all})<0$. Next, we aim for an
asymptotic expansion around $N\to\infty$. For large $N$ we can expand the
exponential function $e^{\alpha/N}= 1+\alpha/N+(\alpha/N)^2/2+\dots$ such that 
\begin{align}\label{eqAsymptoticExpansion}
  \left(1-e^{\alpha/N}\right)^{1/N}
  &= \left(1-1-\alpha/N-(\alpha/N)^2/2+\dots\right)^{1/N}\nonumber\\
  &= (-\alpha/N)^{1/N}\left[1-(\alpha/N)/2+\dots\right]^{1/N}
\end{align}
Restricting ourselves to the leading order we get, 
\begin{equation}
  \wcc^\dagger \simeq 1-\left(-\alpha/N\right)^{1/N} = 1-e^{\frac{1}{N}\ln(-\alpha/N)}.
\end{equation}
Ensuring  that $\lim_{N\to\infty}\frac{1}{N}\ln(-\alpha/N)=0$ according to
l'Hospital rule and expanding $e^{x}\simeq 1+x$, we obtain
\begin{equation}\label{eqAnsatzSolution}
  \wcc^\dagger \simeq \frac{1}{N}\ln\left(\frac{N}{-\ln(p_\mathrm{all})}\right). 
\end{equation}

Equation~\eqref{eqAnsatzSolution} allows to set the probability to transition
away from the state of full network activity. We chose
\begin{equation}
  \wcc^\dagger = \frac{\ln(N)}{N}\quad\text{s.t.}\quad
  1-p_\mathrm{all}=1-e^{-1}\approx0.632,
\end{equation}
such that the state of full network activity is guaranteed to be not absorbing.
Our explicit choice for $\wcc^\dagger$ has several motivations: First, it is the
simplest choice; Second, for $\mtarget=1$ Eq.~\eqref{eqBNfsc} yields
$\wcc(m=1,N-1)=1-(1-\frac{N-1}{N})^{1/(N-1)}=1-\exp\left[\frac{1}{N-1}\ln(1/N)\right]\approx\ln(N)/N$
such that $\wcc^\dagger\approx\wcc(m=1,N-1)$; and last, $\wcc^\dagger$ highlights
the interpretation in terms of a random Erd\H{o}s-R{\'e}nyi network, namely that
in the limit $N\to\infty$ weights with probability $\ln(N)/N$ set us at the
transition between disconnected graphs ($p_\mathrm{all}=0$) and fully connected
graphs ($p_\mathrm{all}=1$).

\begin{figure*}[t]
  \includegraphics{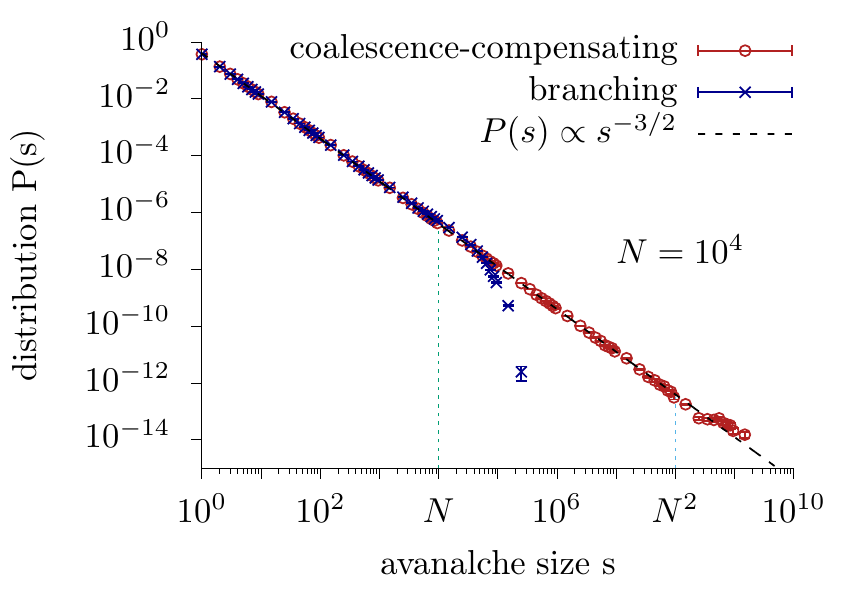}
  \includegraphics{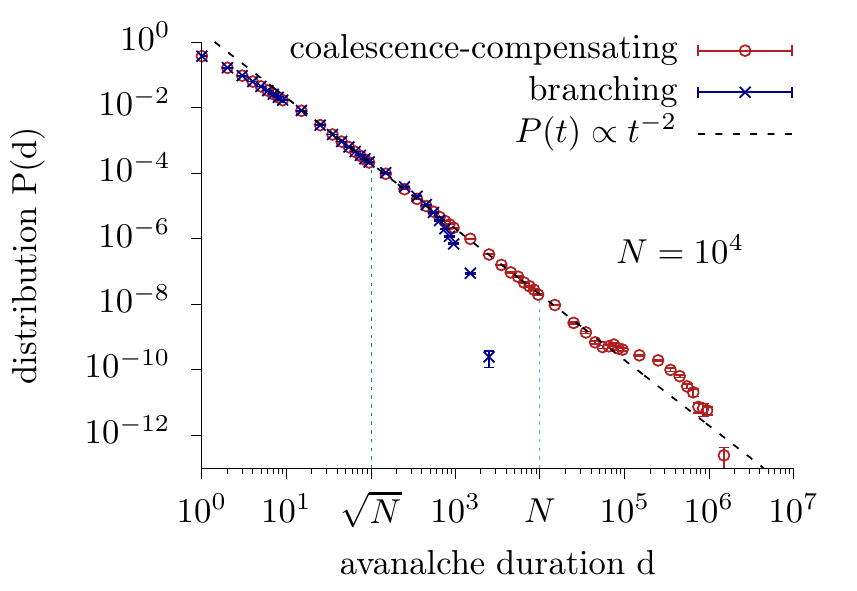}
  \caption{
    Avalanche-size and avalanche-duration distributions for a branching network
    and a coalescence-compensating network with critical dynamics ($m=1$) in the
    separation-of-timescale regime. Compensating coalescence extends the
    power-law behavior of the avalanche-size distribution from $s\sim N$ to
    $s\sim N^2$ and the power-law behavior of the avalanche-duration
    distribution from $d\sim\sqrt{N}$ to $d\sim N$.
    \label{figCCBranchingStatistics}
  }
\end{figure*}
\subsection{S2. Adaptive weights extend the range of critical branching statistics}
\label{secResultCCavalanches}
Above, we argued that compensating for internal coalescence will extend the
range over which a finite network will reproduce statistics of a true branching
process.  In the following, we will show that this is indeed the case even for
critical dynamics ($\mtarget=1$).

We here complement Fig.~\ref{figCorrections} ($N=100$) in the main text by large
networks with $N=10^4$ neurons and critical dynamics ($\mtarget=1$) in the
separation-of-timescale regime (Fig.~\ref{figCCBranchingStatistics}). Compared
to a branching network of the same size, the avalanche-size distribution and
avalanche-duration distributions in the coalescence-compensating model show a
drastically extended power law. In particular, it appears that the power-law
regime is extended from avalanche sizes $s\sim N$ to $s\sim N^2$ and from
avalanche duration $d\sim\sqrt{N}$ to $d\sim N$. The results are preserved in
the driven regime with small external input rate $h\ll1$ (not shown).

We can understand the extended range of critical branching statistics by
considering the universal shape of avalanches. It was shown that the average
time development $A(t,d)$ of neural avalanches with the same duration $d$
collapses on a universal shape $\mathcal{F}(x)$, if properly rescaled as
$A(t,d)=d\mathcal{F}(t/d)$ for our model~\cite{friedman2012}. It can be
anticipated that the power-law characteristics in finite-size branching networks
extends up to $s_\mathrm{max}=\mathcal{O}(N)$. Considering any reasonable
parabolic avalanche shape $\mathcal{F}(x)$ between rectangular and triangular,
the area always scales as $s\sim d\, A^\mathrm{peak}$ where the peak itself
scales as $A^\mathrm{peak}\sim d$ (according to the universal avalanche shape
where $\mathcal{F}^\mathrm{peak}=\text{const}$) such that $s\sim d^2$.  With a
maximal avalanche size $s_\mathrm{max}=\mathcal{O}(N)$, we thus expect a maximal
avalanche duration $d_\mathrm{max}=\mathcal{O}(\sqrt{N})$. This expectation
nicely agrees with our numerical observation
(Fig.~\ref{figCCBranchingStatistics} and main Fig.~\ref{figCorrections}). In
particular, this means that -- due to coalescence -- typical avalanches of
duration $d_\mathrm{max}$ have a maximum number of simultaneously activated
neurons that scales as $A^\mathrm{peak}_\mathrm{max}\sim
d_\mathrm{max}\sim\sqrt{N}$.

Compensating coalescence now shifts the potential maximum number of
simultaneously activated neurons from $A_\mathrm{max}^\mathrm{peak}\sim\sqrt{N}$ to
$A_\mathrm{max}^\mathrm{peak}\sim N$. Because of the universal avalanche shape,
the maximum duration consequently scales as $d_\mathrm{max}\sim
A_\mathrm{max}^\mathrm{peak}\sim N$ and the maximal size then scales as
$s_\mathrm{max}\sim d_\mathrm{max}A_\mathrm{max}^\mathrm{peak}\sim N^2$.

%
\subsection{S3. Adaptive-weight networks have dynamic range and discriminable intervals with properties from both branching process and branching network} 
\label{secResultCCdynamicrange}
If compensating for coalescence extends the range of avalanche statistics, one
could conclude, that the resulting coalescence-compensating model is ``more
critical'' than the branching network. Intuitively, one may expect that this
leads to more optimal information processing and other benefits that come with
operating with critical dynamics. For example, one could expect that the dynamic
range, which is maximal for critical-like dynamics~\cite{kinouchi2006}, is even
larger in coalescence-compensating networks with critical dynamics. In the
following, we will show that on the level of a single network, the dynamic range
is not improved when we compensate for coalescence. 


To analytically calculate the dynamic range of the branching network, we start
with the mean-field approximation Eq.~\eqref{eqMFapproximation} and solve for
the external input
\begin{equation}
  h\bn(m,A) = -\frac{1}{\Delta
  t}\ln\left[\frac{1-\frac{A}{N}}{\left(1-\frac{m}{N}\right)^A}\right].
\end{equation}
For sufficiently large $N$, we approximate \mbox{$(1-m/N)^A\to e^{-ma}$} with
neuron rate $a=A/N$ and find the system-size independent result 
\begin{equation}\label{eqDRh}
  h\bn(m,a) = -\frac{1}{\Delta t}\ln\left[(1-a)e^{ma}\right].
\end{equation}
Recalling our result for the system-size independent neuron rate $a\bn(m,h)$,
Eq.~\eqref{eqResponseSTD}, we continue with the bounds
$a_\mathrm{min}=a(m,h\to0)=1+W\left[-me^{-m}\right]/m$ and
$a_\mathrm{max}=a(m,h\to\infty)=1+W[0]/m=1$ such that $a_x = 1 +
(1-x)W\left[-me^{-m}\right]/m$. Inserting this into Eq.~\eqref{eqDRh}, we obtain
\begin{align}
  h\bn(m,a_x) &= -\frac{1}{\Delta t}\ln\left[(1-x)e^{-xW(-me^{-m})}\right]\\
           &=\frac{xW(-me^{-m})-\ln(1-x)}{\Delta t}.
\end{align}
With this we can calculate the dynamic range of the branching network
\begin{equation}
  \Delta\bn(m) =
  10\log_{10}\left[\frac{0.9W(-me^{-m})-\ln(0.1)}{0.1W(-me^{-m})-\ln(0.9)}\right].
\end{equation}

\begin{figure}
  \includegraphics{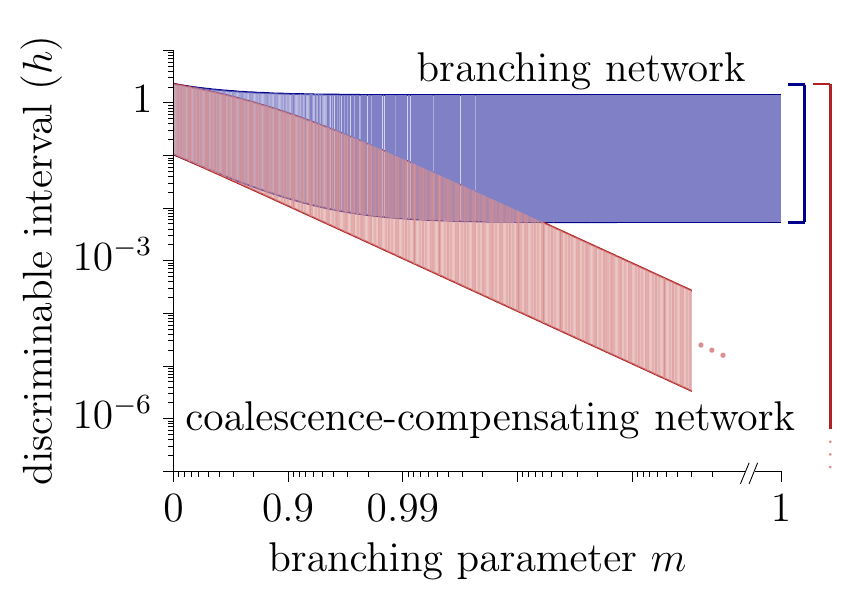}
  \caption{%
    Influence of compensating coalescence on the discriminable interval. For the
    branching network, the discriminable interval barely changes especially for
    branching parameters close to critical-like dynamics ($m=1$). For the
    coalescence-compensating network, the discriminable interval becomes a
    function of the branching parameter.
    \label{figResultDiscriminableInterval}
  }
\end{figure}

For the coalescence-compensating network, we first need to calculate the neuron
rate as response to external input rate. Assuming stationary activity
$A\approx\langle\langle A_{t+1}|A_t\rangle\rangle$, we can use
Eq.~\eqref{eqBNcc2} to write down the mean-field approximation
\begin{equation}
  A\cc= \mtarget[1-\lambda(h)]A\cc+N\lambda(h).
\end{equation}
Solving this for the neuron rate, we obtain 
\begin{equation}\label{eqDRfscResponse}
  a\cc(\mtarget,h) = \frac{\lambda(h)}{1-\mtarget(1-\lambda(h))}.
\end{equation}
This rate is only finite and non-negative for $m<1$. In this range,
$a_\mathrm{min}=0$ and $a_\mathrm{max}=1$ always, such that $a_x = x$.
Calculating the inverse of Eq.~\eqref{eqDRfscResponse},
\begin{equation}
  h\cc(\mtarget,a) = -\frac{1}{\Delta t}\ln\left[ 1-\frac{(1-\mtarget)a}{1-\mtarget
  a}\right],
\end{equation}
we find the dynamic range for coalescence-compensating networks ($\mtarget<1$)
\begin{equation}
  \Delta\cc(m) = 10\log_{10}\left[
              \frac{\ln\left(1-\frac{(1-\mtarget)0.9}{1-\mtarget 0.9}\right)}
              {\ln\left(1-\frac{(1-\mtarget)0.1}{1-\mtarget
              0.1}\right)}\right].
\end{equation}
For coalescence-compensating networks the interval of discriminable input rates
is a function of the branching parameter
(Fig.~\ref{figResultDiscriminableInterval}). 

The results for the coalescence-compensating network are consistent with a
modified branching process. Consider a branching process with upper-bound
population activity $A_t\leq N$ and external input per time step $Nh\Delta t$.
The stationary activity is $A\bp(m,h)=Nh\Delta t/(1-m)$~\cite{harris1963}. The
discriminating activity is defined as $A_x = xN$ such that the inverse of the
activity yields
\begin{equation}
  h\bp(A_x)=(1-m)x/\Delta t,
\end{equation}
and the dynamic range is independent of the branching parameter
\begin{equation}
  \Delta\bp(m) = 10\log_{10}(0.9/0.1) \approx 9.5.
\end{equation}
Importantly, the discriminable interval
$[h\bp(A_{0.1}),h\bp(A_{0.9})]$ is highly depending on $m$.  


\end{document}